\documentclass[onecolumn]{IEEEtran}

\usepackage[ruled]{algorithm2e}
\renewcommand{\algorithmcfname}{ALGORITHM}
\SetAlFnt{\small}
\SetAlCapFnt{\small}
\SetAlCapNameFnt{\small}
\SetAlCapHSkip{0pt}
\IncMargin{-\parindent}
\usepackage{graphicx}
\usepackage{algorithmic}
\usepackage{subfig}
\usepackage{multirow}
\usepackage{array}
\usepackage{amsmath} 
\usepackage{amssymb}
\usepackage[maxfloats=30]{morefloats}
\usepackage{amsthm}
\theoremstyle{definition}

\theoremstyle{theorem}
\newtheorem{theorem}{Theorem}[section]
\begin{document}
		\title{\huge Stable Desynchronization for Wireless Sensor Networks: \\(III) Stability Analysis}

\author{
	\IEEEauthorblockN{Supasate Choochaisri, Kittipat Apicharttrisorn, Chalermek Intanagonwiwat}\\
	\IEEEauthorblockA{Chulalongkorn University, Bangkok, Thailand\\Email:\{supasate.c, kittipat.api, intanago\}@gmail.com}
}

	
	\date{}
	\maketitle
\begin{abstract}
In this paper, we use dynamical systems to analyze stability of desynchronization algorithms at equilibrium. We start by illustrating the equilibrium of a dynamic systems and formalizing force components and time phases. Then, we use Linear Approximation to obtain Jaconian ($J$) matrixes which are used to find the eigenvalues. Next, we employ the Hirst and Macey theorem \cite{hirst97} and Gershgorins theorem \cite{gersheng} to find the bounds of those eigenvalues. Finally, if the number of nodes ($n$) is within such bounds, the systems are stable at equilibrium. (This paper is the last part of the series Stable Desynchronization for Wireless Sensor Networks - (I) Concepts and Algorithms (II) Performance Evaluation (III) Stability Analysis)
\end{abstract}

\section{Stability Analysis of DWARF (the Single-Hop Desynchronization Algorithm)}
To prove that the system is stable, we begin by transforming the system into a non-linear dynamic system. 
The evenness of the number of nodes affects the analysis. Therefore, we divide the non-linear dynamic system into two cases: when $n$ is even and when $n$ is odd where $n$ is the number of nodes.

\begin{figure*}[h]
\centerline{
	\subfloat[]{\includegraphics[scale=0.30]{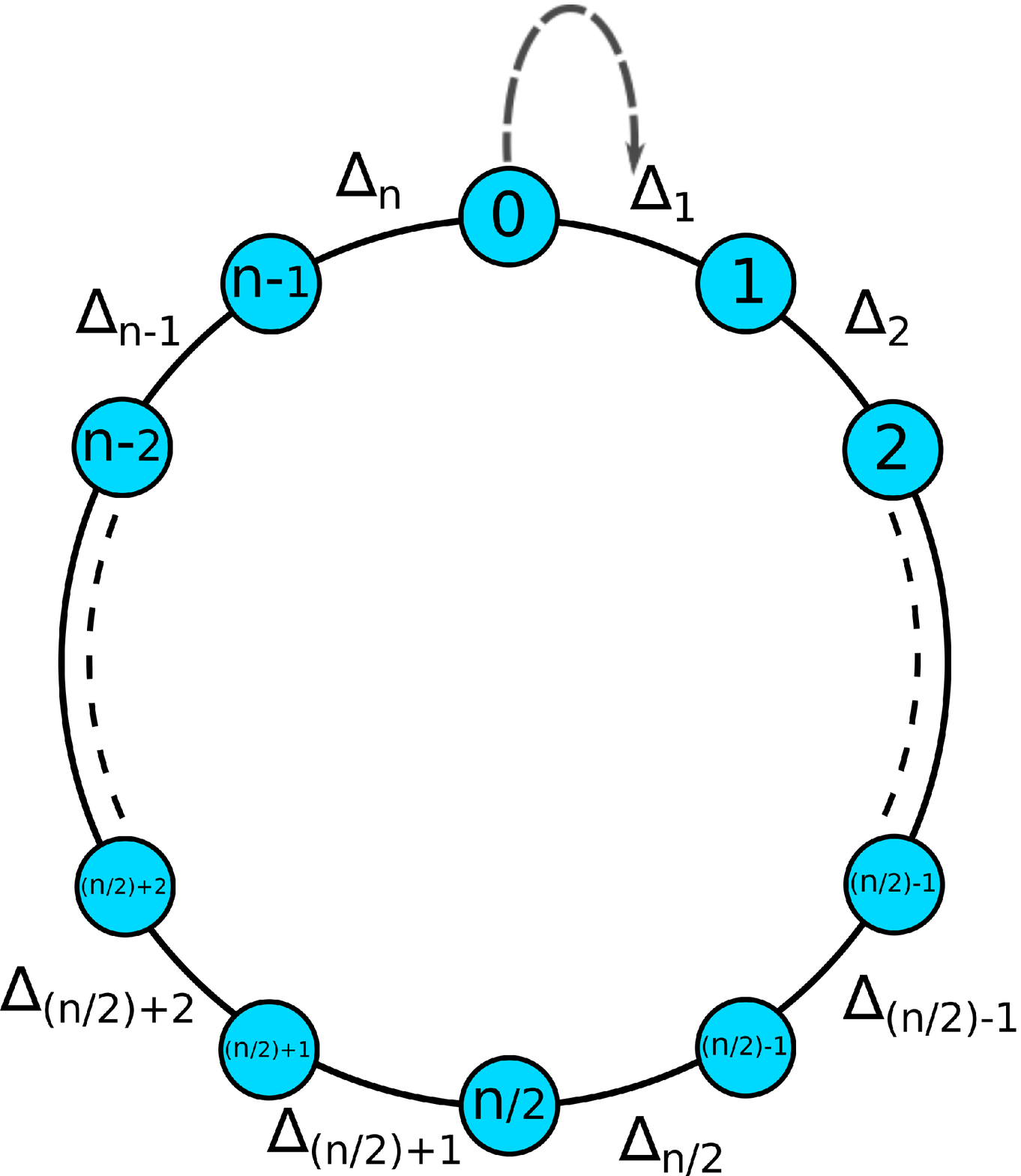}%
	\label{fig:even-perfect}}
	\hfil
	\subfloat[]{\includegraphics[scale=0.30]{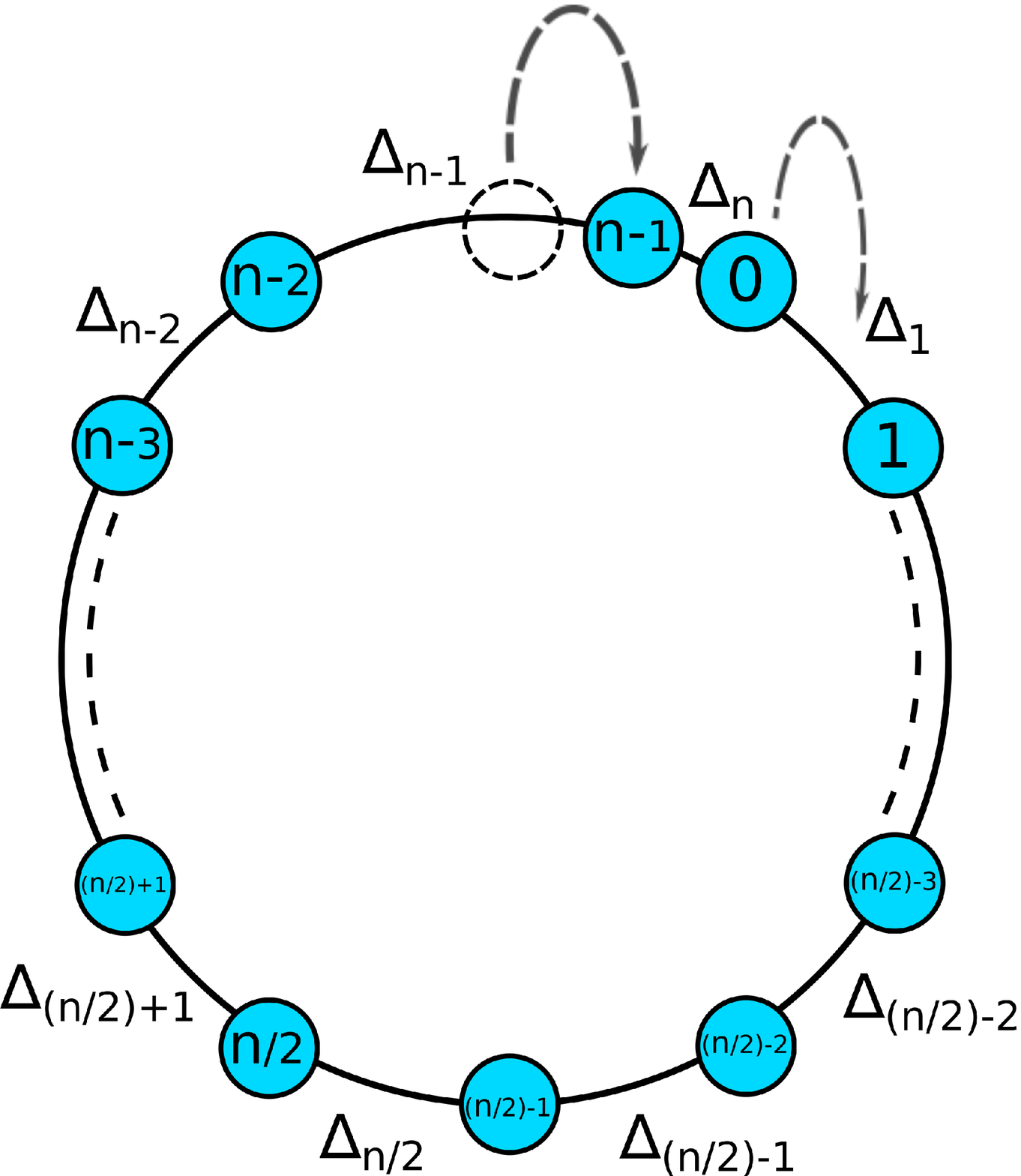}%
	\label{fig:even-adapt}}
}
\caption{Non-linear dynamic system when $n$ is even. (a) The snapshot of the system at one time step. (b) In next time step after node 0 adjusts its phase, node 1 in previous round is re-labelled to 0, node 2 is re-labelled to 1, and so on.}
\label{fig:n-even}
\end{figure*}

1) when $n$ is even: Figure \ref{fig:even-perfect} illustrates the equilibrium of a dynamic system when the number of nodes is even. Noticeably, node 0 and $n/2$ are exactly at the opposite side of each other. At the first snapshot of the system, node 0 adjusts its phase based on the force function of DWARF. After adjustment, we re-label node 1 to 0, node 2 to 1, ..., node $n-1$ to $n-2$, and node 0 to $n-1$ for analysis at the next snapshot (see Figure \ref{fig:even-adapt}).
Therefore, to transform into a difference equation, $\Delta_{1}$ in the next snapshot is $\Delta_{2}$ in the previous snapshot, $\Delta_{2}$ in the next snapshot is $\Delta_{3}$ in the previous snapshot, and so on. However, $\Delta_{n-1}$ and $\Delta_{n}$ in the next snapshot are $\Delta_{n}$ and $\Delta_{1}$ in the previous snapshot adjusted by the force function, respectively.

Therefore, the non-linear dynamic system when $n$ is even can be expressed as the following difference equations:

\begin{alignat}{1}
 \Delta_{1} =& \Delta_{2} \nonumber \\
 \Delta_{2} =& \Delta_{3} \nonumber \\
 \vdots \nonumber \\
 \Delta_{n-2} =& \Delta_{n-1} \nonumber  \\
 \Delta_{n-1} =& \Delta_{n} + KT\Bigg( -\frac{1}{\Delta_{1}} - \frac{1}{\Delta_{1} + \Delta_{2}} - \hdots - \frac{1}{\Delta_{1} + \Delta_{2} + \hdots + \Delta_{\frac{n}{2}-1}} \nonumber \\ 
&+ \frac{1}{\Delta_{n}} + \frac{1}{\Delta_{n} + \Delta_{n-1}} + \hdots + \frac{1}{\Delta_{n} + \Delta_{n-1} + \hdots + \Delta_{\frac{n}{2}+2}}\Bigg) \nonumber  \\
 \Delta_{n} =& \Delta_{1} - KT\Bigg( -\frac{1}{\Delta_{1}} - \frac{1}{\Delta_{1} + \Delta_{2}} - \hdots - \frac{1}{\Delta_{1} + \Delta_{2} + \hdots + \Delta_{\frac{n}{2}-1}} \nonumber \\ 
&+ \frac{1}{\Delta_{n}} + \frac{1}{\Delta_{n} + \Delta_{n-1}} + \hdots + \frac{1}{\Delta_{n} + \Delta_{n-1} + \hdots + \Delta_{\frac{n}{2}+2}}\Bigg) 
\end{alignat}

We note that the force from node $n/2$ in the previous snapshot is already balanced when we consider node 0. Therefore, $\Delta_{\frac{n}{2}}$ and $\Delta_{\frac{n}{2}+1}$ do not appear in the force equation for adjusting $\Delta_{n-1}$ and $\Delta_{n}$ in the next snapshot.

\begin{figure*}[h]
\centerline{
	\subfloat[]{\includegraphics[scale=0.30]{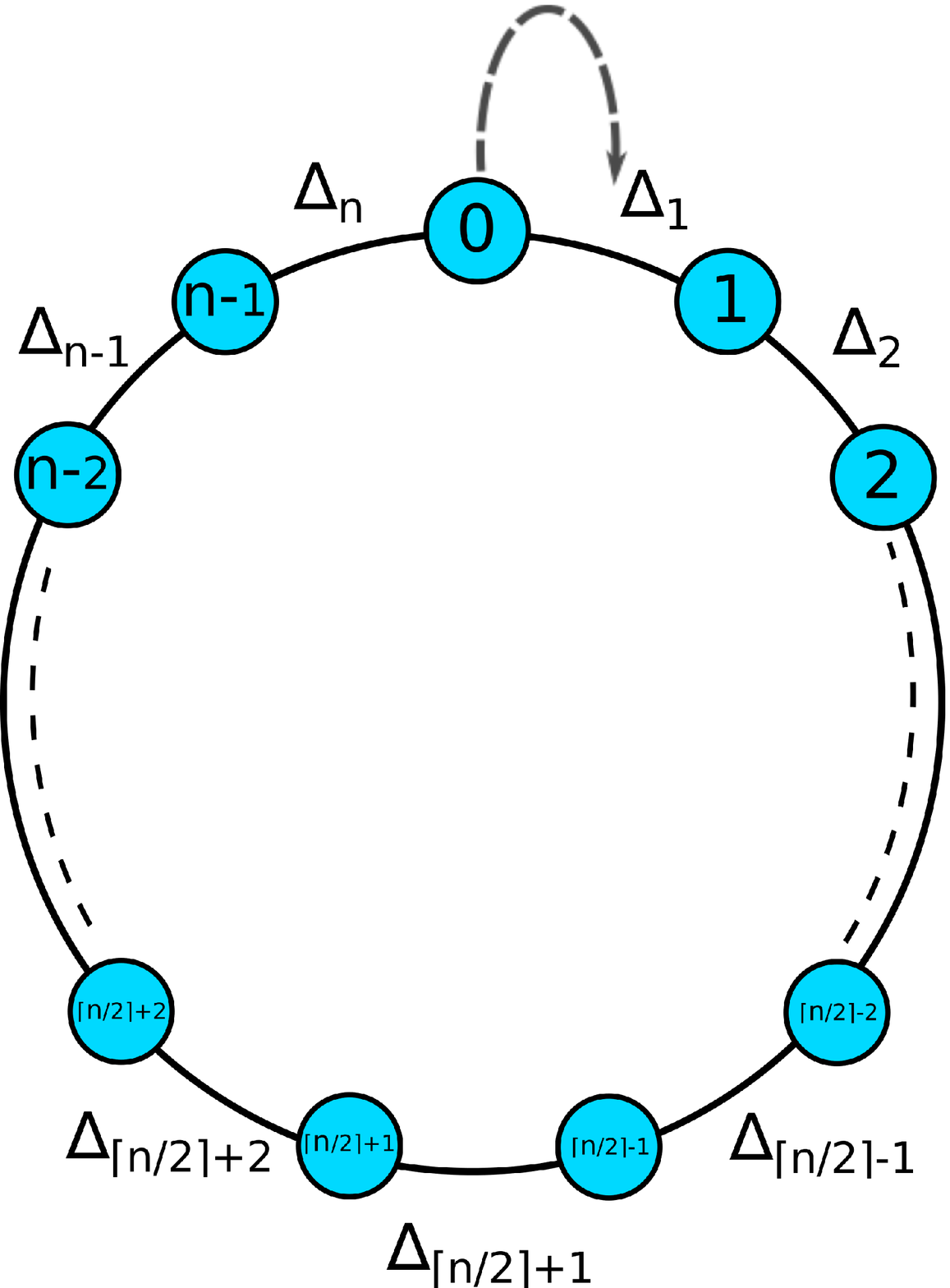}%
	\label{fig:odd}}
	\hfil
	\subfloat[]{\includegraphics[scale=0.30]{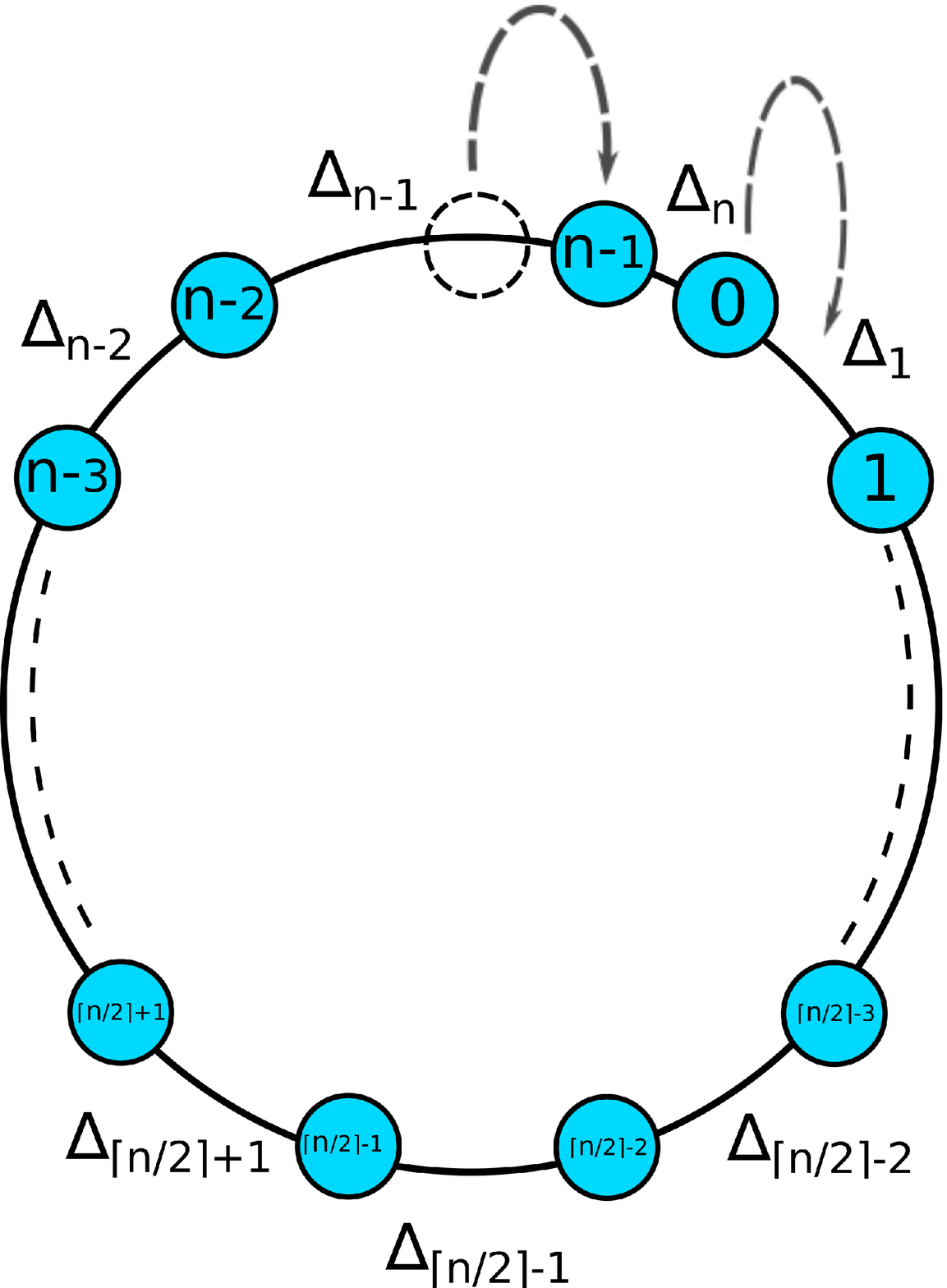}%
	\label{fig:odd-adapt}}
}
\caption{Non-linear dynamic system when $n$ is odd. (a) The snapshot of the system at one time step. (b) In next time step after node 0 adjusts its phase, node 1 in previous round is re-labelled to 0, node 2 is re-labelled to 1, and so on.}
\label{fig:n-odd}
\end{figure*}

2) when $n$ is odd:
Similarly, Figure \ref{fig:n-odd} illustrates the non-linear dynamic system when $n$ is odd. The only difference from when $n$ is even is that there is no node that is opposite to node 0. Therefore, in the difference equations, only $\Delta_{\lceil\frac{n}{2}\rceil}$ from the previous snapshot does not appear in the force equation for adjusting $\Delta_{n-1}$ and $\Delta_{n}$ in the next snapshot.

\begin{alignat}{1}
 \Delta_{1} =& \Delta_{2} \nonumber \\
 \Delta_{2} =& \Delta_{3} \nonumber \\
 \vdots \nonumber  \\
 \Delta_{n-2} =& \Delta_{n-1} \nonumber \\
 \Delta_{n-1} =& \Delta_{n} + KT\Bigg( -\frac{1}{\Delta_{1}} - \frac{1}{\Delta_{1} + \Delta_{2}} - \hdots - \frac{1}{\Delta_{1} + \Delta_{2} + \hdots + \Delta_{\lceil{\frac{n}{2}}\rceil-1}} \nonumber \\ 
&+ \frac{1}{\Delta_{n}} + \frac{1}{\Delta_{n} + \Delta_{n-1}} + \hdots + \frac{1}{\Delta_{n} + \Delta_{n-1} + \hdots + \Delta_{\lceil\frac{n}{2}\rceil+1}}\Bigg) \nonumber \\
 \Delta_{n} =& \Delta_{1} - KT\Bigg( -\frac{1}{\Delta_{1}} - \frac{1}{\Delta_{1} + \Delta_{2}} - \hdots - \frac{1}{\Delta_{1} + \Delta_{2} + \hdots + \Delta_{\lceil\frac{n}{2}\rceil-1}} \nonumber \\ 
&+ \frac{1}{\Delta_{n}} + \frac{1}{\Delta_{n} + \Delta_{n-1}} + \hdots + \frac{1}{\Delta_{n} + \Delta_{n-1} + \hdots + \Delta_{\lceil\frac{n}{2}\rceil+1}}\Bigg) \nonumber 
\end{alignat}

Due to the non-linear adaptation function, the standard linear dynamic system analysis does not suffice.
Therefore, we locally analyse stability of the system around a fixed point which is the equilibrium point.
We begin by linear approximation to find the Jacobian at the equilibrium. Then, from the Jacobian, we find the bound of eigenvalues which is the crucial part of stability analysis.

\subsection{Linear Approximation}
The Jacobian ($J$) of a difference equations system is defined as follows:
\begin{alignat}{2}
J=\begin{pmatrix} 
{\frac{\partial \Delta_{1}}{\partial \Delta_{1}}}  & {\frac{\partial \Delta_{1}}{\partial \Delta_{2}}} & \cdots & {\frac{\partial \Delta_{1}}{\partial \Delta_{n-1}}} & {\frac{\partial \Delta_{1}}{\partial \Delta_{n}}} \\ 
{\frac{\partial \Delta_{2}}{\partial \Delta_{1}}}  & {\frac{\partial \Delta_{2}}{\partial \Delta_{2}}} & \cdots &  {\frac{\partial \Delta_{2}}{\partial \Delta_{n-1}}} & {\frac{\partial \Delta_{2}}{\partial \Delta_{n}}} \\  
\vdots & \vdots & \ddots & \vdots & \vdots \\
{\frac{\partial \Delta_{n-1}}{\partial \Delta_{1}}} & {\frac{\partial \Delta_{n-1}}{\partial \Delta_{2}}} & \cdots &  {\frac{\partial \Delta_{n-1}}{\partial \Delta_{n-1}}} & {\frac{\partial \Delta_{n-1}}{\partial \Delta_{n}}} \\  
{\frac{\partial \Delta_{n}}{\partial \Delta_{1}}} & {\frac{\partial \Delta_{n}}{\partial \Delta_{2}}} & \cdots &  {\frac{\partial \Delta_{n}}{\partial \Delta_{n-1}}} & {\frac{\partial \Delta_{n}}{\partial \Delta_{n}}} \\  
\end{pmatrix}
\end{alignat}

We consider the Jacobian when $n$ is even and odd separately. In both cases, after finding the Jacobian, we substitute each $\Delta_{i}$ with $T/n$ which is the phase interval between each node at the equilibrium.

\subsubsection{ $n$ is even} 
For $\Delta_{i}, 1 \leq i \leq n-2$, partial derivatives of ${\partial \Delta_{i} / \partial \Delta_{j}}, 1 \leq j \leq n$ are the followings:
\begin{alignat}{1}
  {\frac{\partial \Delta_{i}}{\partial \Delta_{j}}} = \left\{ 
  \begin{array}{l l}
    1 & \quad \text{if } j = i + 1\\
    0 & \quad \text{otherwise}\\
  \end{array} \right. \nonumber
\end{alignat}

For $\Delta_{n-1}$, we find its partial derivatives as follows:
\begin{alignat}{1}
{\frac{\partial \Delta_{n-1}}{\partial \Delta_{1}}} =& KT\left(\frac{1}{\Delta_1^2} + \frac{1}{(\Delta_1 + \Delta_2)^2} + \cdots + \frac{1}{(\Delta_1 + \Delta_2 + \cdots + \Delta_{\frac{n}{2}-1})^2}\right) \nonumber \\
=&KT\left(\frac{n^2}{T^2}\right)\left(\frac{1}{1^2} + \frac{1}{2^2} + \cdots + \frac{1}{(\frac{n}{2}-1)^2}\right) = \frac{Kn^2}{T} \sum\limits_{i=1}^{\frac{n}{2}-1}{\frac{1}{i^2}} \nonumber \\
{\frac{\partial \Delta_{n-1}}{\partial \Delta_{2}}} =& KT\left(\frac{1}{(\Delta_1 + \Delta_2)^2} + \cdots + \frac{1}{(\Delta_1 + \Delta_2 + \cdots + \Delta_{\frac{n}{2}-1})^2}\right) \nonumber \\
=&KT\left(\frac{n^2}{T^2}\right)\left(\frac{1}{2^2} + \cdots + \frac{1}{(\frac{n}{2}-1)^2}\right) = \frac{Kn^2}{T} \sum\limits_{i=2}^{\frac{n}{2}-1}{\frac{1}{i^2}} \nonumber \\
\vdots \nonumber \\
{\frac{\partial \Delta_{n-1}}{\partial \Delta_{\frac{n}{2}-1}}} =& KT\left(\frac{1}{(\Delta_1 + \Delta_2 + \cdots + \Delta_{\frac{n}{2}-1})^2}\right) \nonumber \\
=&KT\left(\frac{n^2}{T^2}\right)\left(\frac{1}{(\frac{n}{2}-1)^2}\right) = \frac{Kn^2}{T} \sum\limits_{i=\frac{n}{2}-1}^{\frac{n}{2}-1}{\frac{1}{i^2}} \nonumber \\	
{\frac{\partial \Delta_{n-1}}{\partial \Delta_{\frac{n}{2}}}} =& 0 \nonumber \\
{\frac{\partial \Delta_{n-1}}{\partial \Delta_{\frac{n}{2}+1}}} =& 0 \nonumber \\
{\frac{\partial \Delta_{n-1}}{\partial \Delta_{\frac{n}{2}+2}}} =& KT\left(-\frac{1}{(\Delta_n + \Delta_{n-1} + \cdots + \Delta_{\frac{n}{2}+2})^2}\right) \nonumber \\
=&-KT\left(\frac{n^2}{T^2}\right)\left(\frac{1}{(\frac{n}{2}-1)^2}\right) = -\frac{Kn^2}{T} \sum\limits_{i=\frac{n}{2}-1}^{\frac{n}{2}-1}{\frac{1}{i^2}} \nonumber \\	
\vdots \nonumber \\
{\frac{\partial \Delta_{n-1}}{\partial \Delta_{n-1}}} =& KT\left(-\frac{1}{(\Delta_n + \Delta_{n-1})^2} - \cdots - \frac{1}{(\Delta_n + \Delta_{n-1} + \cdots + \Delta_{\frac{n}{2}+2})^2}\right) \nonumber \\
=&-KT\left(\frac{n^2}{T^2}\right)\left(\frac{1}{2^2} + \cdots + \frac{1}{(\frac{n}{2}-1)^2}\right) = -\frac{Kn^2}{T} \sum\limits_{i=2}^{\frac{n}{2}-1}{\frac{1}{i^2}} \nonumber \\
{\frac{\partial \Delta_{n-1}}{\partial \Delta_{n}}} =& 1+KT\left(-\frac{1}{\Delta_n^2} - \frac{1}{(\Delta_n + \Delta_{n-1})^2} - \cdots - \frac{1}{(\Delta_n + \Delta_{n-1} + \cdots + \Delta_{\frac{n}{2}+2})^2}\right) \nonumber  \\
=&1-KT\left(\frac{n^2}{T^2}\right)\left(\frac{1}{1^2} + \frac{1}{2^2} + \cdots + \frac{1}{(\frac{n}{2}-1)^2}\right) = 1-\frac{Kn^2}{T} \sum\limits_{i=1}^{\frac{n}{2}-1}{\frac{1}{i^2}} \nonumber 
\end{alignat}

For $\Delta_{n}$, we find its partial derivatives which is similar to $\Delta_{n-1}$ as follows.

\begin{alignat}{1}
{\frac{\partial \Delta_{n}}{\partial \Delta_{1}}} =& 1-\frac{Kn^2}{T} \sum\limits_{i=1}^{\frac{n}{2}-1}{\frac{1}{i^2}} \nonumber \\
{\frac{\partial \Delta_{n}}{\partial \Delta_{2}}} =& -\frac{Kn^2}{T} \sum\limits_{i=2}^{\frac{n}{2}-1}{\frac{1}{i^2}} \nonumber \\
\vdots \nonumber \\
{\frac{\partial \Delta_{n}}{\partial \Delta_{\frac{n}{2}-1}}} =& -\frac{Kn^2}{T} \sum\limits_{i=\frac{n}{2}-1}^{\frac{n}{2}-1}{\frac{1}{i^2}} \nonumber \\	
{\frac{\partial \Delta_{n}}{\partial \Delta_{\frac{n}{2}}}} =& 0 \nonumber \\
{\frac{\partial \Delta_{n}}{\partial \Delta_{\frac{n}{2}+1}}} =& 0 \nonumber \\
{\frac{\partial \Delta_{n}}{\partial \Delta_{\frac{n}{2}+2}}} =& \frac{Kn^2}{T} \sum\limits_{i=\frac{n}{2}-1}^{\frac{n}{2}-1}{\frac{1}{i^2}} \nonumber \\	
\vdots \nonumber \\
{\frac{\partial \Delta_{n}}{\partial \Delta_{n-1}}} =& \frac{Kn^2}{T} \sum\limits_{i=2}^{\frac{n}{2}-1}{\frac{1}{i^2}} \nonumber \\
{\frac{\partial \Delta_{n}}{\partial \Delta_{n}}} =& \frac{Kn^2}{T} \sum\limits_{i=1}^{\frac{n}{2}-1}{\frac{1}{i^2}} \nonumber 
\end{alignat}

Let $\sum\limits_{s}$ stands for $\sum\limits_{i=s}^{\frac{n}{2}-1}{\frac{1}{i^2}}$, where $s \in [1,\frac{n}{2}-1]$ and let $A = Kn^2/T$.
The Jacobian matrix of the system, when $n$ is even, is

\begin{alignat}{1}
J=\begin{pmatrix}
 0 & 1  & \cdots & 0 & 0 & 0 & 0 & \cdots & 0 & 0\\ 
 0 & 0  & \ddots & 0 & 0 & 0 & 0 & \cdots & 0 & 0\\  
\vdots & \vdots & \ddots & \ddots & \vdots & \vdots & \vdots & \ddots & \vdots & \vdots\\
 0 & 0  & \cdots & 0 & 1 & 0 & 0 & \cdots & 0 & 0\\  
 0 & 0  & \cdots & 0 & 0 & 1 & 0 & \cdots & 0 & 0\\  
 0 & 0 & \cdots & 0 & 0 & 0 & 1 & \cdots & 0 & 0\\  
\vdots & \vdots & \ddots & \vdots & \vdots & \vdots & \vdots & \ddots & \vdots & \vdots\\
 0 & 0  & \cdots & 0 & 0 & 0 & 0 & \cdots & 1 & 0 \\ 
A\sum\limits_{1} & A\sum\limits_{2} & \cdots &  A\sum\limits_{\frac{n}{2}-1} & 0 & 0 & - A\sum\limits_{\frac{n}{2}-1}& \cdots & - A\sum\limits_{2} & 1 - A\sum\limits_{1} \\  
1 - A\sum\limits_{1} & - A\sum\limits_{2} & \cdots &  - A\sum\limits_{\frac{n}{2}-1} & 0 & 0 &  A\sum\limits_{\frac{n}{2}-1}& \cdots & A\sum\limits_{2} & A\sum\limits_{1} \\  
\end{pmatrix}.
\end{alignat}

\subsubsection{ $n$ is odd} 
Similarly, when $n$ is odd, we can find its Jacobian with the same procedure as when $n$ is even.We obtain the similar Jacobian matrix as follows: 

\begin{alignat}{1}
J=\begin{pmatrix}
 0 & 1  & \cdots & 0 & 0 & 0 & \cdots & 0 & 0\\ 
 0 & 0  & \ddots & 0 & 0 & 0 & \cdots & 0 & 0\\  
\vdots & \vdots & \ddots & \ddots & \vdots & \vdots & \ddots & \vdots & \vdots\\
 0 & 0  & \cdots & 0 & 1  & 0 & \cdots & 0 & 0\\  
 0 & 0  & \cdots & 0 & 0 & 1 & \cdots & 0 & 0\\  
\vdots & \vdots & \ddots & \vdots & \vdots & \vdots & \ddots & \vdots & \vdots\\
 0 & 0  & \cdots & 0 & 0 & 0 & \cdots & 1 & 0 \\ 
A\sum\limits_{1} & A\sum\limits_{2} & \cdots &  A\sum\limits_{\frac{n}{2}-1} & 0 & - A\sum\limits_{\frac{n}{2}-1}& \cdots & - A\sum\limits_{2} & 1 - A\sum\limits_{1} \\  
1 - A\sum\limits_{1} & - A\sum\limits_{2} & \cdots &  - A\sum\limits_{\frac{n}{2}-1} & 0 &  A\sum\limits_{\frac{n}{2}-1}& \cdots & A\sum\limits_{2} & A\sum\limits_{1} \\  
\end{pmatrix}.
\end{alignat}

\subsection{Finding Eigenvalues}

After finding the Jacobian of the linear approximation at the equilibrium, we find the eigenvalues by solving the equation $|J-\lambda I| = 0$. If all eigenvalues lay on a unit circle, the system is stable at the equilibrium. Therefore, we begin by finding the determinant of the matrix $J-\lambda I$. 
We use row operations to transform the determinant of $|J-\lambda I|$ into the determinant of a triangular matrix. Then, the determinant of the triangular matrix is the multiplication of diagonal entries. 

The following procedure is to transform the determinant of $|J-\lambda I|$ into the determinant of triangular matrix when $n$ is even:

\begingroup
\small
\begin{alignat}{1}
|J-\lambda I| &=\begin{vmatrix} 
-\lambda  & 1 & \cdots & 0 & 0 & 0 & 0 & \cdots & 0 & 0\\ 
0  & -\lambda  & \ddots & 0 & 0 & 0 & 0 & \cdots & 0 & 0\\  
\vdots & \vdots & \ddots & \ddots & \vdots & \vdots & \vdots & \ddots & \vdots & \vdots\\
0  & 0  & \cdots & -\lambda & 1 & 0 & 0 & \cdots & 0 & 0\\  
0  & 0  & \cdots & 0 & -\lambda & 1 & 0 & \cdots & 0 & 0\\  
0  & 0  & \cdots & 0 & 0 & -\lambda & 1 & \cdots & 0 & 0\\  
\vdots & \vdots & \ddots & \vdots & \vdots & \vdots & \ddots & \ddots & \vdots & \vdots\\
0  & 0  & \cdots & 0 & 0 & 0 & 0 & \cdots & 1 & 0 \\ 
A\sum\limits_{1} & A\sum\limits_{2} & \cdots &  A\sum\limits_{\frac{n}{2}-1} & 0 & 0 & -A\sum\limits_{\frac{n}{2}-1}& \cdots & -A\sum\limits_{2}-\lambda & 1-A\sum\limits_{1} \\  
1-A\sum\limits_{1} & -A\sum\limits_{2} & \cdots &  -A\sum\limits_{\frac{n}{2}-1} & 0 & 0 &  A\sum\limits_{\frac{n}{2}-1}& \cdots & A\sum\limits_{2} & A\sum\limits_{1}-\lambda \\  
\end{vmatrix}  \\
&=\begin{vmatrix} 
-\lambda  & 1  & \cdots & 0 & 0 & 0 & 0 & \cdots & 0 & 0\\ 
0  & -\lambda  & \ddots & 0 & 0 & 0 & 0 & \cdots & 0 & 0\\  
\vdots & \vdots & \ddots & \ddots & \vdots & \vdots & \vdots & \ddots & \vdots & \vdots\\
0  & 0  & \cdots & -\lambda & 1 & 0 & 0 & \cdots & 0 & 0\\  
0  & 0  & \cdots & 0 & -\lambda & 1 & 0 & \cdots & 0 & 0\\  
0  & 0  & \cdots & 0 & 0 & -\lambda & 1 & \cdots & 0 & 0\\  
\vdots & \vdots & \ddots & \vdots & \vdots & \vdots & \ddots & \ddots & \vdots & \vdots\\
0  & 0  & \cdots & 0 & 0 & 0 & 0 & \cdots & 1 & 0 \\ 
1 & 0 & \cdots &  0 & 0 & 0 & 0 & \cdots & -\lambda & 1-\lambda \\  
1-A\sum\limits_{1} & -A\sum\limits_{2} & \cdots &  -A\sum\limits_{\frac{n}{2}-1} & 0 & 0 &  A\sum\limits_{\frac{n}{2}-1}& \cdots & A\sum\limits_{2} & A\sum\limits_{1}-\lambda \\  
\end{vmatrix} \displaybreak \\
&=\begin{vmatrix} 
-\lambda + \frac{1}{\lambda^{n-2}}  & 0  & \cdots & 0 & 0 & 0 & 0 & \cdots & 0 &  \frac{1-\lambda}{\lambda^{n-2}}\\ 
\frac{1}{\lambda^{n-3}}  & -\lambda  & \cdots & 0 & 0 & 0 & 0 & \cdots & 0 &  \frac{1-\lambda}{\lambda^{n-3}} \\ 
\frac{1}{\lambda^{n-4}}  & 0  & \ddots & 0 & 0 & 0 & 0 & \cdots & 0 &  \frac{1-\lambda}{\lambda^{n-4}} \\  
\vdots & \vdots & \ddots & \ddots & \vdots & \vdots & \vdots & \ddots & \vdots & \vdots\\
\frac{1}{\lambda^{\frac{n}{2}-2}}  & 0  & \cdots & 0 & -\lambda & 0 & 0 & \cdots & 0 & \frac{1-\lambda}{\lambda^{\frac{n}{2}-2}}\\  
\frac{1}{\lambda^{\frac{n}{2}-3}}  & 0  & \cdots & 0 & 0 & -\lambda & 0 & \cdots & 0 & \frac{1-\lambda}{\lambda^{\frac{n}{2}-3}}\\  
\vdots & \vdots & \ddots & \vdots & \vdots & \vdots & \ddots & \ddots & \vdots & \vdots\\
 \frac{1}{\lambda}  & 0  & \cdots & 0 & 0 & 0 & 0 & \ddots & 0 &  \frac{1-\lambda}{\lambda}\\ 
1 & 0 & \cdots &  0 & 0 & 0 & 0 & \cdots & -\lambda & 1-\lambda \\  
1-A\sum\limits_{1} & -A\sum\limits_{2} & \cdots &  -A\sum\limits_{\frac{n}{2}-1} & 0 & 0 &  A\sum\limits_{\frac{n}{2}-1}& \cdots & A\sum\limits_{2} & A\sum\limits_{1}-\lambda \\ 
\end{vmatrix}  \\
&=\begin{vmatrix} 
-\lambda + \frac{1}{\lambda^{n-2}}  & 0  & \cdots & 0 & 0 & 0 & 0 & \cdots & 0 &  \frac{1-\lambda}{\lambda^{n-2}}\\ 
\lambda^{2} & -\lambda  & \cdots & 0 & 0 & 0 & 0 & \cdots & 0 &  0 \\ 
\lambda^{3} & 0  & \ddots & 0 & 0 & 0 & 0 & \cdots & 0 &  0 \\  
\vdots & \vdots & \ddots & \ddots & \vdots & \vdots & \vdots & \ddots & \vdots & \vdots\\
\lambda^{\frac{n}{2}}  & 0  & \cdots & 0 & -\lambda & 0 & 0 & \cdots & 0 & 0 \\  
\lambda^{\frac{n}{2}+1}  & 0  & \cdots & 0 & 0 & -\lambda & 0 & \cdots & 0 & 0 \\  
\vdots & \vdots & \ddots & \vdots & \vdots & \vdots & \ddots & \ddots & \vdots & \vdots\\
\lambda^{n-2}  & 0  & \cdots & 0 & 0 & 0 & 0 & \ddots & 0 & 0 \\ 
\lambda^{n-1}  & 0 & \cdots &  0 & 0 & 0 & 0 & \cdots & -\lambda & 0 \\  
1-A\sum\limits_{1} & -A\sum\limits_{2} & \cdots &  -A\sum\limits_{\frac{n}{2}-1} & 0 & 0 &  A\sum\limits_{\frac{n}{2}-1}& \cdots & A\sum\limits_{2} & A\sum\limits_{1}-\lambda \\  
\end{vmatrix} \\
&=\begin{vmatrix} 
-\lambda + \frac{1}{\lambda^{n-2}}  & 0  & \cdots & 0 & 0 & 0 & 0 & \cdots & 0 &  \frac{1-\lambda}{\lambda^{n-2}}\\ 
\lambda^{2} & -\lambda  & \cdots & 0 & 0 & 0 & 0 & \cdots & 0 &  0 \\ 
\lambda^{3} & 0  & \ddots & 0 & 0 & 0 & 0 & \cdots & 0 &  0 \\  
\vdots & \vdots & \ddots & \ddots & \vdots & \vdots & \vdots & \ddots & \vdots & \vdots\\
\lambda^{\frac{n}{2}}  & 0  & \cdots & 0 & -\lambda & 0 & 0 & \cdots & 0 & 0 \\  
\lambda^{\frac{n}{2}+1}  & 0  & \cdots & 0 & 0 & -\lambda & 0 & \cdots & 0 & 0 \\  
\vdots & \vdots & \ddots & \vdots & \vdots & \vdots & \ddots & \ddots & \vdots & \vdots\\
\lambda^{n-2}  & 0  & \cdots & 0 & 0 & 0 & 0 & \ddots & 0 & 0 \\ 
\lambda^{n-1}  & 0 & \cdots &  0 & 0 & 0 & 0 & \cdots & -\lambda & 0 \\  
B & 0 & \cdots &  0 & 0 & 0 & 0 & \cdots & 0 & A\sum\limits_{1}-\lambda \\  
\end{vmatrix} \\ 
&=\begin{vmatrix} 
-\lambda + \frac{1}{\lambda^{n-2}} + \frac{B}{A\sum\limits_{1}-\lambda}\left(\frac{\lambda-1}{\lambda^{n-2}}\right)  & 0  & \cdots & 0 & 0 & 0 & 0 & \cdots & 0 & 0\\ 
\lambda^{2} & -\lambda  & \cdots & 0 & 0 & 0 & 0 & \cdots & 0 &  0 \\ 
\lambda^{3} & 0  & \ddots & 0 & 0 & 0 & 0 & \cdots & 0 &  0 \\  
\vdots & \vdots & \ddots & \ddots & \vdots & \vdots & \vdots & \ddots & \vdots & \vdots\\
\lambda^{\frac{n}{2}}  & 0  & \cdots & 0 & -\lambda & 0 & 0 & \cdots & 0 & 0 \\  
\lambda^{\frac{n}{2}+1}  & 0  & \cdots & 0 & 0 & -\lambda & 0 & \cdots & 0 & 0 \\  
\vdots & \vdots & \ddots & \vdots & \vdots & \vdots & \ddots & \ddots & \vdots & \vdots\\
\lambda^{n-2}  & 0  & \cdots & 0 & 0 & 0 & 0 & \ddots & 0 & 0 \\ 
\lambda^{n-1}  & 0 & \cdots &  0 & 0 & 0 & 0 & \cdots & -\lambda & 0 \\  
B & 0 & \cdots &  0 & 0 & 0 & 0 & \cdots & 0 & A\sum\limits_{1}-\lambda \\  
\end{vmatrix}
\end{alignat}
\endgroup

where 
\begin{equation}
\label{eq:deteven}
B = 1-A\sum\limits_{1} - \lambda A\sum\limits_{2} - \lambda^{2}A\sum\limits_{3} - \cdots - \lambda^{\frac{n}{2}-2} A\sum\limits_{\frac{n}{2}-1} + \lambda^{\frac{n}{2}+1} A\sum\limits_{\frac{n}{2}-1} + \cdots + \lambda^{n-2}A\sum\limits_2. 
\end{equation}

Similarly, we use the same procedure to find the determinant of $|J-\lambda I|$ when $n$ is odd. The result matrix is the same as above except that

\begin{equation}
\label{eq:detodd}
B = 1-A\sum\limits_{1} - \lambda A\sum\limits_{2} - \lambda^{2}A\sum\limits_{3} - \cdots - \lambda^{\lceil\frac{n}{2}\rceil-1} A\sum\limits_{\frac{n}{2}-1} + \lambda^{\lceil\frac{n}{2}\rceil+1} A\sum\limits_{\frac{n}{2}-1} + \cdots + \lambda^{n-2}A\sum\limits_2. 
\end{equation}

We note that the difference between Equation \ref{eq:deteven} and \ref{eq:detodd} is the exponents of $\lambda$ at two middle terms.

The determinant of a triangular matrix is the product of diagonal terms. That is
\begin{equation}
|J-\lambda I| = \left(-\lambda + \frac{1}{\lambda^{n-2}} + \frac{B}{A\sum\limits_{1}-\lambda}\left(\frac{\lambda-1}{\lambda^{n-2}}\right)\right)(-\lambda)(-\lambda)\cdots(-\lambda)(-\lambda)\left(A\sum\limits_{1}-\lambda\right) = 0. \nonumber
\end{equation}

For terms $-\lambda$, it is obvious that we have repeated eigenvalues $\lambda$ that are equal to $0$ and lay on a unit circle.

For the last term $A\sum\limits_{1}-\lambda$, the value of $\lambda$ depends on $n$ as follows:
\begin{alignat}{1}
A\sum\limits_{1}-\lambda &= 0 \nonumber \\
\lambda &= A\sum\limits_{1} \nonumber \\
\lambda &= 0.038597n^{0.126}\sum\limits_{i=1}^{\frac{n}{2}-1}{\frac{1}{i^2}} \geq 0.\nonumber 
\end{alignat}
In this case, $\lambda \geq 0$, however, we have to find the least upper limit of $n$ that leads to $\lambda \leq 1$. 
\begin{alignat}{2}
\lambda = 0.038597n^{0.126}\sum\limits_{i=1}^{\frac{n}{2}-1}{\frac{1}{i^2}} \leq & 1 \nonumber  \\
0.038597n^{0.126}(1.645) \leq & 1 \nonumber \\
n \leq & 3.178 \times 10^9.
\end{alignat}
Therefore, if the number of nodes $n$ is less than $3.178 \times 10^9$ nodes, the eigenvalue $\lambda$ lies between 0 and 1 in this case.

For the first term, $-\lambda + \frac{1}{\lambda^{n-2}} + \frac{B}{A\sum\limits_{1}-\lambda}\left(\frac{\lambda-1}{\lambda^{n-2}}\right)$, the $\lambda$ must not be $0$. If $n$ is even, we derive the polynomial of $\lambda$ as follows:
\begin{alignat}{2}
0 =& -\lambda + \frac{1}{\lambda^{n-2}} + \frac{B}{A\sum\limits_{1}-\lambda}\left(\frac{\lambda-1}{\lambda^{n-2}}\right)  \nonumber \\
0 =& -\lambda^{n-1} + 1 + \frac{B}{A\sum\limits_{1}-\lambda}\left(\lambda-1\right)  \nonumber \\ 
0 =& \lambda^{n} - \lambda^{n-1}A\sum\limits_{1} - \lambda + A\sum\limits_{1} + B(\lambda - 1)   \nonumber  \\ 
0 =& \lambda^{n} - \lambda^{n-1}A\sum\limits_{1} - \lambda + A\sum\limits_{1} & \nonumber \\ 
&+ \lambda -\lambda A\sum\limits_{1} - \lambda^{2}A\sum\limits_{2} - \lambda^{3}A\sum\limits_{3} - \cdots - \lambda^{\frac{n}{2}-1} A\sum\limits_{\frac{n}{2}-1} & \nonumber \\ 
&+ \lambda^{\frac{n}{2}+2} A\sum\limits_{\frac{n}{2}-1} + \cdots  + \lambda^{n-2}A\sum\limits_3 + \lambda^{n-1}A\sum\limits_2   \nonumber \\ 
& - 1 + A\sum\limits_{1} + \lambda A\sum\limits_{2} + \lambda^{2}A\sum\limits_{3} + \cdots + \lambda^{\frac{n}{2}-2} A\sum\limits_{\frac{n}{2}-1} \nonumber \\
& - \lambda^{\frac{n}{2}+1} A\sum\limits_{\frac{n}{2}-1} - \cdots - \lambda^{n-2}A\sum\limits_2 \nonumber \displaybreak \\
0 =& \lambda^{n} - A\lambda^{n-1}\left(\sum\limits_{1} - \sum\limits_{2}\right) - A\lambda^{n-2}\left(\sum\limits_{2} - \sum\limits_{3}\right) - \cdots - A\lambda^{\frac{n}{2}+2}\left(\sum\limits_{\frac{n}{2}-2} - \sum\limits_{\frac{n}{2}-1}\right) \nonumber \\
& - A\lambda^{\frac{n}{2}+1}\sum\limits_{\frac{n}{2}-1} - A\lambda^{\frac{n}{2}-1}\sum\limits_{\frac{n}{2}-1} - A\lambda^{\frac{n}{2}-2}\left(\sum\limits_{\frac{n}{2}-2} - \sum\limits_{\frac{n}{2}-1}\right) - \cdots - A\lambda^{2}\left(\sum\limits_{2} - \sum\limits_{3}\right) \nonumber \\
& - A\lambda\left(\sum\limits_{1} - \sum\limits_{2}\right) + 2A\sum\limits_{1} - 1\nonumber \\
0 =& \lambda^{n} - \frac{A\lambda^{n-1}}{1^2} - \frac{A\lambda^{n-2}}{2^2} - \cdots - \frac{A\lambda^{\frac{n}{2}+2}}{(\frac{n}{2}-2)^2} - \frac{A\lambda^{\frac{n}{2}+1}}{(\frac{n}{2}-1)^2} \nonumber \\
& - \frac{A\lambda^{\frac{n}{2}-1}}{(\frac{n}{2}-1)^2} - \frac{A\lambda^{\frac{n}{2}-2}}{(\frac{n}{2}-2)^2}  - \cdots - \frac{A\lambda^{2}}{2^2}  - \frac{A\lambda}{1^2} + 2A\sum\limits_{1} - 1\nonumber 
\end{alignat}

To get the eigenvalues, we have to solve the polynomial
\begin{equation}
f_{even}(\lambda) = \lambda^{n} - \frac{A\lambda^{n-1}}{1^2} - \frac{A\lambda^{n-2}}{2^2} - \cdots  - \frac{A\lambda^{\frac{n}{2}+1}}{(\frac{n}{2}-1)^2} - \frac{A\lambda^{\frac{n}{2}-1}}{(\frac{n}{2}-1)^2} - \cdots - \frac{A\lambda^{2}}{2^2}  - \frac{A\lambda}{1^2} + 2A\sum\limits_{1} - 1 = 0.
\label{eq:even_polynomial}
\end{equation}

Similarly, if $n$ is odd, we have to solve the polynomial
\begin{equation}
f_{odd}(\lambda) = \lambda^{n} - \frac{A\lambda^{n-1}}{1^2} - \frac{A\lambda^{n-2}}{2^2} - \cdots  - \frac{A\lambda^{\lceil\frac{n}{2}\rceil}}{(\frac{n}{2}-1)^2} - \frac{A\lambda^{\lceil\frac{n}{2}\rceil-1}}{(\frac{n}{2}-1)^2} - \cdots - \frac{A\lambda^{2}}{2^2}  - \frac{A\lambda}{1^2} + 2A\sum\limits_{1} - 1 = 0.
\label{eq:odd_polynomial}
\end{equation}

However, according to the Abel's Impossibility theorem (or Abel-Ruffini theorem), we cannot find a general algebraic solution to polynomial equations of degree five or higher (\cite{abel}). Therefore, instead of finding the exact values, we find the upper and lower bound of the eigenvalues to ensure that they lay on a unit circle.

\subsection{The Bound of Eigenvalues}
\label{sec:bound}
To find the bound of polynomial roots (\textit{i.e.} our eigenvalues), we follow the following theorem of Hirst and Macey (\cite{hirst97}).
\begin{theorem}[Hirst and Macey Bound]
Given $f: \mathbb{C} \rightarrow \mathbb{C}$ defined by $f(z) = z^n + a_{n-1}z^{n-1} + \hdots + a_{1}z + a_0$, where $a_0, a_1, \hdots , a_n \in \mathcal{C}$, and $n$ a positive integer. If $z$ is a zero of $f$, then
\begin{alignat}{2}
|z| \leq \max \left\{1,\sum_{i=0}^{n-1}|a_i|\right\}. \nonumber
\end{alignat}
\end{theorem}

Therefore, to bound all eigenvalues within a unit circle, $\sum_{i=0}^{n-1}|a_i|$ must be less than or equal 1.
From both Equation \ref{eq:even_polynomial} and \ref{eq:odd_polynomial}, 
\begin{alignat}{2}
\sum_{i=0}^{n-1}|a_i| = \Bigg|2A\sum_{i=1}^{\frac{n}{2}-1}\frac{1}{i^2} - 1\Bigg| + 2\Bigg(\frac{A}{1^2} + \frac{A}{2^2} + \cdots + \frac{A}{(\frac{n}{2}-1)^2}\Bigg) & \leq 1 \nonumber \\
\Bigg|2A\sum_{i=1}^{\frac{n}{2}-1}\frac{1}{i^2} - 1\Bigg| + 2A\sum_{i=1}^{\frac{n}{2}-1}\frac{1}{i^2} & \leq 1 \nonumber \\
\Bigg|2A\sum_{i=1}^{\frac{n}{2}-1}\frac{1}{i^2} - 1\Bigg| & \leq 1 - 2A\sum_{i=1}^{\frac{n}{2}-1}\frac{1}{i^2} \nonumber \\
- 1 + 2A\sum_{i=1}^{\frac{n}{2}-1}\frac{1}{i^2} \leq 2A\sum_{i=1}^{\frac{n}{2}-1}\frac{1}{i^2} - 1 & \leq 1 - 2A\sum_{i=1}^{\frac{n}{2}-1}\frac{1}{i^2},
\end{alignat}
where $A = Kn^2/T = 38.597n^{-1.874}(T/1000)(n^2/T)= 0.038597n^{0.126}$.

For the first case, $- 1 + 2A\sum_{i=1}^{\frac{n}{2}-1}\frac{1}{i^2} \leq 2A\sum_{i=1}^{\frac{n}{2}-1}\frac{1}{i^2} - 1$ is always true regardless of the number of nodes $n$. Therefore, we consider the second case:
\begin{alignat}{2}
2A\sum_{i=1}^{\frac{n}{2}-1}\frac{1}{i^2} - 1 & \leq 1 - 2A\sum_{i=1}^{\frac{n}{2}-1}\frac{1}{i^2} \nonumber \\
4A\sum_{i=1}^{\frac{n}{2}-1}\frac{1}{i^2}  & \leq 2 \nonumber \\
0.038597n^{0.126}\sum_{i=1}^{\frac{n}{2}-1}\frac{1}{i^2}  & \leq \frac{1}{2}. 
\label{eq:maceybound}
\end{alignat}

When $n$ is large, $\sum_{i=1}^{\frac{n}{2}-1}\frac{1}{i^2}$ is approximately equal to the Reimann zeta function $\zeta (2) = \sum_{i=1}^{\infty}\frac{1}{i^2} = \pi^2/6 \approx 1.645$. Therefore, we get the following:
\begin{alignat}{2}
0.038597n^{0.126}(1.645)  & \leq \frac{1}{2} \nonumber \\
n^{0.126} &\leq 7.87499981 \nonumber \\
n &\leq 1.29 \times 10^7.
\end{alignat}

\section{Stability Analysis of M-DWARF (the Multi-Hop Desynchronization Algorithm)}

The multi-hop stability analysis in this chapter is more complicated than the analysis of single-hop networks due to the connectivity and topology which affect the analysis.
We begin by formalizing components of forces to a general form. Then, we transform the system into a dynamic system equations and locally analyse at the equilibrium.

\subsection{Definition}
We first define several variables and notations to be used throughout the analysis in this section.

We define $n$ to be the number of nodes in the system and $T$ be the time period. We assume $n$ is even. However, in the case of odd $n$, the system can be analysed with the same procedure.

We define $c_{i,j}$ to be a connectivity status in two-hop communication from node $j$ to node $i$.
If node $i$ perceives the presence of node $j$ (as a one-hop neighbor or by relaying relative phase), $c_{i,j}$ is 1. Otherwise, $c_{i,j}$ is 0. For example, in a simple 3-node chain network with node labelled 0, 1, and 2 respectively, if communication links are symmetry, the values of $c_{0,1}, c_{1,0}, c_{1,2}$, and $c_{2,1}$ are 1 whereas the values of $c_{0,2}$ and $c_{2,0}$ are 0. We assume that nodes are labelled sequentially in increasing order from $0$ to $n-1$ according the the increasing phase in the global period ring.

We define the modulo notation $[x]_n$ to stand for $x \mod n$ for brevity.

We define $\Delta_i$ to be a phase interval between node $i$ and node $[i+1]_n$, where $\Delta_i \in (0, T]$.

\subsection{Force Component}
To construct a dynamic system, we first analyse how force with the absorption mechanism affects the system equations.
As described in Section 4.2.3, a force can be absorbed if it is not originated from the closest phase neighbor. Therefore, respect to the node $i$'s point of view, some forces have no effect to node $i$ whereas some forces do. 
We classify forces that affect node $i$ into three components: closest component, resistance component, and absorption component.
The final form of forces that affect node $i$ will be in the following form:
\begin{alignat}{2}
F_{i} =& (\text{closest component} + \text{resistance component} - \text{absorption component})_{positive} \nonumber \\
&- (\text{closest component} + \text{resistance component} - \text{absorption component})_{negative}.
\end{alignat}

\textit{Closest component}:
Respect to the node $i$'s point of view, the closest component is the force from the closest phase neighbor of node $i$ and node $i$ perceives its presence. This force is not absorbed by any node. This force component from node $j$ to node $i$ will appear in the equation if, between node $j$ and node $i$, node $i$ does not perceive the presence of any node. In other words, node $j$ is the closest perceived phase neighbor of node $i$. We firstly define the closest component for node $i$ by using the combination of logic and algebraic expression as follows (we use it only for clarification purpose and we will change it to pure algebraic expression later),
\begin{alignat}{1}
& \Bigg((c_{i,[i+(n-1)]_n})f_{i,[i+(n-1)]_n}^{+} + (\lnot c_{i,[i+(n-1)]_n} \land c_{i,[i+(n-2)]_n})f_{i,[i+(n-2)]_n}^{+} \nonumber \\
&+ (\neg c_{i,[i+(n-1)]_n}\land \neg c_{i,[i+(n-2)]_n} \land c_{i,[i+(n-3)]_n})f_{i,[i+(n-3)]_n}^{+} + \cdots  \nonumber \\
&+ (\neg c_{i,[i+(n-1)]_n}\land \neg c_{i,[i+(n-2)]_n} \land \cdots \land \neg c_{i,[i+(\frac{n}{2}+2)]_n} \land c_{i,[i+(\frac{n}{2}+1)]_n})f_{i, [i + (\frac{n}{2}+1)]_n}^{+}\Bigg) \nonumber \\
&-\Bigg((c_{i,[i+1]_n})f_{i,[i+1]_n}^{-} + (\lnot c_{i,[i+1]_n} \land c_{i,[i+2]_n})f_{i,[i+2]_n}^{-} \nonumber \\
&+ (\neg c_{i,[i+1]_n}\land \neg c_{i,[i+2]_n} \land c_{i,[i+3]_n})f_{i,[i+3]_n}^{-} + \cdots \nonumber \\
&+ (\neg c_{i,[i+1]_n}\land \neg c_{i,[i+2]_n} \land \cdots \land \neg c_{i,[i+(\frac{n}{2}-2)]_n} \land c_{i,[i+(\frac{n}{2}-1)]_n})f_{i, [i + (\frac{n}{2}-1)]_n}^{-}\Bigg),\nonumber \\
\label{eq:closest}
\end{alignat}
where $f_{i,[j]_n}^{+}$ is a positive (clockwise) force and $f_{i,[j]_n}^{-}$ is a negative (counter-clockwise) force.
We note that $f_{i,[j]_n}^{+}$ appears only when $c_{i,[j]_n} = 1$ and all $c_{i,[k]_n} = 0$, where  $j< k < i + n$. Similary, $f_{i,[j]_n}^{-}$ appears only when $c_{i,[j]_n} = 1$ and all $c_{i,[k]_n} = 0$, where  $i< k < j$.

Then, we convert the logic expression into the algebraic expression. For \textit{logical negation}, $\neg c_{i,[j]_n}$ can be algebraically expressed as $(1 - c_{i,[j]_n})$. For \textit{logical and} ($\land$), we can express algebraically by using multiplication instead. For example, $c_{i,[j]_n} \land c_{i,[k]_n}$ can be expressed as $c_{i,[j]_n}c_{i,[k]_n}$. 

Therefore, Equation \ref{eq:closest} can be expressed algebraically as the following,
\begin{alignat}{2}
&\Bigg((c_{i,[i+(n-1)]_n})f_{i,[i+(n-1)]_n}^{+} + ((1 - c_{i,[i+(n-1)]_n})c_{i,[i+(n-2)]_n})f_{i,[i+(n-2)]_n}^{+} \nonumber \\
&+ ((1 - c_{i,[i+(n-1)]_n})(1 -  c_{i,[i+(n-2)]_n})c_{i,[i+(n-3)]_n})f_{i,[i+(n-3)]_n}^{+} + \cdots  \nonumber \\
&+ ((1 - c_{i,[i+(n-1)]_n})((1- c_{i,[i+(n-2)]_n}) \cdots (1 - c_{i,[i+(\frac{n}{2}+2)]_n})c_{i,[i+(\frac{n}{2}+1)]_n})f_{i, [i + (\frac{n}{2}+1)]_n}^{+}\Bigg) \nonumber \\
&-\Bigg((c_{i,[i+1]_n})f_{i,[i+1]_n}^{-} + ((1 - c_{i,[i+1]_n})c_{i,[i+2]_n})f_{i,[i+2]_n}^{-} \nonumber \\
&+ ((1 - c_{i,[i+1]_n})(1 -  c_{i,[i+2]_n})c_{i,[i+3]_n})f_{i,[i+3]_n}^{-} + \cdots  \nonumber \\
&+ ((1 - c_{i,[i+1]_n})((1- c_{i,[i+2]_n}) \cdots (1 - c_{i,[i+(\frac{n}{2}-2)]_n})c_{i,[i+(\frac{n}{2}-1)]_n})f_{i, [i + (\frac{n}{2}-1)]_n}^{-}\Bigg) \nonumber \\
&= \sum_{j=i+(\frac{n}{2}+1)}^{i + (n-1)}f_{i,[j]_n}^{+}c_{i,[j]_n}\prod_{k=1}^{i + n - (j + 1)}(1-c_{i,[i+(n-k)]_n}) - \sum_{j=i+1}^{i + (\frac{n}{2}-1)}f_{i,[j]_n}^{-}c_{i,[j]_n}\prod_{k=1}^{j-1}(1-c_{i,[i+k]_n}).
\end{alignat}

Let $R_{i,j}^{+}$ be $\prod_{k=1}^{i + n - (j + 1)}(1-c_{i,[i+(n-k)]_n}) \in {0,1}$ and $R_{i,j}^{-}$ be $\prod_{k=1}^{j-1}(1-c_{i,[i+k]_n}) \in {0,1}$. We derive the following form of the closet component to be used in our analysis,
\begin{alignat}{2}
\sum_{j=i+(\frac{n}{2}+1)}^{i + (n-1)}f_{i,[j]_n}^{+}c_{i,[j]_n}R_{i,j}^{+} - \sum_{j=i+1}^{i + (\frac{n}{2}-1)}f_{i,[j]_n}^{-}c_{i,[j]_n}R_{i,j}^{-}.
\end{alignat}

\textit{Resistance component}:
Respect to the node $i$'s point of view, there is a resistance component originated from node $j$ if the following criteria are satisfied:
\begin{itemize}
	\item Node $i$ perceives the presence of node $j$.
	\item There is at least one node $k$ following node $j$ in the time period ring (in each force direction). 
	\item Node $i$ perceives the presence of node $k$.
\end{itemize}
For example, with respect to the node $0$'s point of view, if node $1$ is the closest phase neighbor of node $0$ and there is node $2$ following node $1$ in clockwise direction, and node $0$ perceives the presence of both node $1$ and node $2$, then, there is a force difference between node $1$ and node $2$ in the form of $f_{0,1} - f_{0,2}$ (see Section 4.2.3). The part $f_{0,1}$ is called \textit{resistance} component and the part $f_{0,2}$ is called \textit{absorption} component which we will describe later. We note that, if there is no node following the closest phase neighbor in each direction, there is no resistance and absorption components.

Therefore, we define the resistance component for node $i$ by using the combination of logic and algebraic expression as follows,

\begin{alignat}{2}
&\Bigg(((c_{i,[i+(n-2)]_n} 
\lor c_{i,[i+(n-3)]_n} 
\lor \cdots \lor c_{i,[i+(\frac{n}{2} + 1)]_n})
\land c_{i,[i+(n-1)]_n}) f_{i,[i+(n-1)]_n}^{+} \nonumber \\
&+ ((c_{i,[i+(n-3)]_n} \lor c_{i,[i+(n-4)]_n} 
\lor \cdots \lor c_{i,[i+(\frac{n}{2} + 1)]_n})
\land c_{i,[i+(n-2)]_n}) f_{i,[i+(n-2)]_n}^{+} \nonumber \\
&+ \cdots + ((c_{i,[i+(\frac{n}{2} + 1)]_n})\land c_{i,[i+(\frac{n}{2}+2)]_n})f_{i,[i+(\frac{n}{2}+2)]_n}^{+}\Bigg) \nonumber \\
&-\Bigg(((c_{i,[i+2]_n} 
\lor c_{i,[i+3]_n} \lor \cdots \lor c_{i,[i+(\frac{n}{2} - 1)]_n})
\land c_{i,[i+1]_n})  f_{i,[i+1]_n}^{-} \nonumber \\
&+ ((c_{i,[i+3]_n} \lor c_{i,[i+4]_n} \lor \cdots \lor c_{i,[i+(\frac{n}{2} - 1)]_n})
\land c_{i,[i+2]_n})  f_{i,[i+2]_n}^{-} \nonumber \\
&+ \cdots + ((c_{i,[i+(\frac{n}{2} - 1)]_n}) \land c_{i,[i+(\frac{n}{2}-2)]_n})f_{i,[i+(\frac{n}{2}-2)]_n}^{-}\Bigg).
\label{eq:logicalor}
\end{alignat}

Then, we convert the logical expression into the algebraic expression.

Let  $s^{+}(v_{i,k}^+)$ be an \textit{algebraic or} function to represent a \textit{logical or} expression of $(c_{i,[i+(n-k)]_n} \lor c_{i,[i+(n-k-1)]_n} \lor \cdots \lor c_{i,[i+(\frac{n}{2}+1)]_n})$, where $v_{i,k}^+ =  c_{i,[i+(\frac{n}{2}+1)]_n}2^{\frac{n}{2}-1-k} + \cdots + c_{i,[i+(n-k-1)]_n}2^{1} + c_{i,[i+(n-k)]_n}2^{0}$. 

Similarly, let $s^{-}(v_{i,k}^-)$ be an algebraic function to represent a \textit{logical or} expression of $(c_{i,[i+k]_n} \lor c_{i,[i+(k+1)]_n} \lor \cdots \lor c_{i,[i+(\frac{n}{2}-1)]_n})$, where $v_{i,k}^- = c_{i,[i+(\frac{n}{2}-1)]_n}2^{\frac{n}{2}-1-k} + \cdots + c_{i,[i+(k+1)]_n}2^{1} + c_{i,[i+k]_n}2^{0}$.

We note that, if we write a binary string $c_{i,[i+(\frac{n}{2}+1)]_n} \cdots c_{i,[i+(n-k-1)]_n}c_{i,[i+(n-k)]_n}$, $v_k^+$ is an integer value in base 10 of this binary string where $v_{i,k}^+ \in \lbrace 0,1,\cdots,2^{\frac{n}{2}-k}-1 \rbrace$. If all $c_{i,[j]_n} = 0$, then $v_{i,k}^+ = 0$. The result is similar for $v_{i,k}^-$.

Let $I_A(x)$ be an indicator function as follows,
\begin{alignat}{2}
I_A(x) = \begin{cases}1 & \text{if }x \in A \\ 0 & \text{if }x \notin A. \end{cases}
\label{eq:indicator}
\end{alignat}

Therefore, $s^+(v_{i,k}^+) = I_{ \mathbb{N} - \lbrace 0 \rbrace}(v_{i,k}^+)$ and $s^-(v_{i,k}^-) = I_{ \mathbb{N} - \lbrace 0 \rbrace}(v_{i,k}^-)$. 

From Equation \ref{eq:logicalor}, we derive the equivalent algebraic expression as follows,
\begin{alignat}{2}
&\Bigg(((s^+(v_{i,2}^+))c_{i,[i+(n-1)]_n}) f_{i,[i+(n-1)]_n}^{+} + ((s^+(v_{i,3}^+))c_{i,[i+(n-2)]_n}) f_{i,[i+(n-2)]_n}^{+} \nonumber \\
&+ \cdots + ((s^+(v_{i,\frac{n}{2} - 1}^+))c_{i,[i+(\frac{n}{2}+2)]_n})f_{i,[i+(\frac{n}{2}+2)]_n}^{+}\Bigg) \nonumber \\
&-\Bigg(((s^-(v_{i,2}^-))c_{i,[i+1]_n})  f_{i,[i+1]_n}^{-} + ((s^-(v_{i,3}^-))c_{i,[i+2]_n})  f_{i,[i+2]_n}^{-} \nonumber \\
&+ \cdots + ((s^-(v_{i,\frac{n}{2} - 1}^-))c_{i,[i+(\frac{n}{2}-2)]_n})f_{i,[i+(\frac{n}{2}-2)]_n}^{-}\Bigg).
\end{alignat}

Let $S_{i,j}^{+} = s^+(v_{i,i+n-j+1}^+) \in \lbrace 0,1 \rbrace$ and $S_{i,j}^{-} = s^-(v_{i,j-i+1}^-) \in \lbrace 0,1 \rbrace$. We derive the following form for the resistance component to be used in our analysis,
\begin{alignat}{2}
\sum_{j=i+(\frac{n}{2}+2)}^{i + (n-1)}f_{i,[j]_n}^{+}c_{i,[j]_n}S_{i,j}^{+} - \sum_{j=i+1}^{i + (\frac{n}{2}-2)}f_{i,[j]_n}^{-}c_{i,[j]_n}S_{i,j}^{-}.
\end{alignat}

\textit{Absorption component}:  
Respect to the node $i$'s point of view, there is an absorption component originated from node $j$ if the following criteria are satisfied:
\begin{itemize}
	\item Node $i$ perceives the presence of node $j$.
	\item There is at least one node $k$ stays between node $i$ and node $j$ in the time period ring (in each force direction). 
	\item Node $i$ perceives the presence of node $k$.
\end{itemize}

Therefore, we define the absorption component for node $i$ by using the combination of logic and algebraic expression as follows,

\begin{alignat}{2}
&\Bigg(((c_{i,[i+(n-1)]_n})
\land c_{i,[i+(n-2)]_n}) f_{i,[i+(n-2)]_n}^{+} \nonumber \\
&+ ((c_{i,[i+(n-1)]_n} \lor c_{i,[i+(n-2)]_n})
\land c_{i,[i+(n-3)]_n}) f_{i,[i+(n-3)]_n}^{+} \nonumber \\
&+ \cdots + ((c_{i,[i+(n-1)]_n} \lor c_{i,[i+(n-2)]_n} \lor \cdots \lor c_{i,[i+(\frac{n}{2} + 2)]_n})\land c_{i,[i+(\frac{n}{2}+1)]_n})f_{i,[i+(\frac{n}{2}+1)]_n}^{+}\Bigg) \nonumber \\
&-\Bigg(((c_{i,[i+1]_n})
\land c_{i,[i+2]_n})  f_{i,[i+2]_n}^{-} \nonumber \\
&+ ((c_{i,[i+1]_n} \lor c_{i,[i+2]_n})
\land c_{i,[i+3]_n})  f_{i,[i+3]_n}^{-} \nonumber \\
&+ \cdots + ((c_{i,[i+1]_n} \lor c_{i,[i+2]_n} \lor \cdots \lor c_{i,[i+(\frac{n}{2} - 2)]_n}) \land c_{i,[i+(\frac{n}{2}-1)]_n})f_{i,[i+(\frac{n}{2}-1)]_n}^{-}\Bigg).
\label{eq:logicalor2}
\end{alignat}

Based on the same procedure deriving the resistance component, we derive the following form for the absorption component to be used in our analysis,
\begin{alignat}{2}
\sum_{j=i+(\frac{n}{2}+1)}^{i + (n-2)}f_{i,[j]_n}^{+}c_{i,[j]_n}T_{i,j}^{+} - \sum_{j=i+2}^{i + (\frac{n}{2}-1)}f_{i,[j]_n}^{-}c_{i,[j]_n}T_{i,j}^{-},
\end{alignat}

where $T_{i,j}^+$ and $T_{i,j}^-$ are the algebraic expressions of the \textit{logical or} terms in Equation \ref{eq:logicalor2}.

Therefore, respect to the node $i$'s point of view, the total force at node $i$ is as follows,
\begin{alignat}{2}
F_i &= \left ( \sum_{j= i + (\frac{n}{2} + 1)}^{i+(n-1)}  f_{i,[j]_n}^{+}c_{i,[j]_n}R_{i,j}^+ + \sum_{j= i + (\frac{n}{2} + 2)}^{i+(n-1)}  f_{i,[j]_n}^{+}c_{i,[j]_n}S_{i,j}^+ - \sum_{j= i + (\frac{n}{2} + 1)}^{i+(n-2)}  f_{i,[j]_n}^{+}c_{i,[j]_n}T_{i,j}^+\right) \nonumber \\
&- \left ( \sum_{j= i + 1}^{i+(\frac{n}{2}-1)}  f_{i,[j]_n}^{-}c_{i,[j]_n}R_{i,j}^- + 
\sum_{j= i + 1}^{i+(\frac{n}{2}-2)}  f_{i,[j]_n}^{-}c_{i,[j]_n}S_{i,j}^- -
\sum_{j= i + 2}^{i+(\frac{n}{2}-1)}  f_{i,[j]_n}^{-}c_{i,[j]_n}T_{i,j}^- \right).
\label{eq:algebraforce}
\end{alignat}

In the next section, we analyse the stability of the M-DWARF algorithm based on the derived total force.

\subsection{Stability Analysis}
As same as the analysis of single-hop networks, to prove that the system is stable, we begin by transforming the system into a non-linear dynamic system. 

Let $f_{i,j}^{+}$ and $f_{i,j}^{-}$ be the positive and negative forces from node $j$ to node $i$ respectively,
\begin{alignat}{2}
f_{i,j}^{+} = \frac{T}{\sum_{k=j}^{i-1}\Delta_{[k]_n}} \text{ and } f_{i,j}^{-} = \frac{T}{\sum_{k=i}^{j-1}\Delta_{[k]_n}},  
\end{alignat}
where $\Delta_k \in (0, T]$ is the phase difference between node $[k + 1]_n$ and node $k$.

From Equation \ref{eq:algebraforce}, the total force at node $i$ is
\begin{alignat}{2}
F_i &= \left ( \sum_{j= i + (\frac{n}{2} + 1)}^{i+(n-1)}  f_{i,[j]_n}^{+}c_{i,[j]_n}R_{i,j}^+ + \sum_{j= i + (\frac{n}{2} + 2)}^{i+(n-1)}  f_{i,[j]_n}^{+}c_{i,[j]_n}S_{i,j}^+ - \sum_{j= i + (\frac{n}{2} + 1)}^{i+(n-2)}  f_{i,[j]_n}^{+}c_{i,[j]_n}T_{i,j}^+\right) \nonumber \\
&- \left ( \sum_{j= i + 1}^{i+(\frac{n}{2}-1)}  f_{i,[j]_n}^{-}c_{i,[j]_n}R_{i,j}^- + 
\sum_{j= i + 1}^{i+(\frac{n}{2}-2)}  f_{i,[j]_n}^{-}c_{i,[j]_n}S_{i,j}^- -
\sum_{j= i + 2}^{i+(\frac{n}{2}-1)}  f_{i,[j]_n}^{-}c_{i,[j]_n}T_{i,j}^- \right).
\end{alignat}

Let $\Delta_i$ be the current phase difference between node $[i+1]_n$ and node $i$ and $\Delta_i'$ be the phase difference between node $[i+1]_n$ and node $i$ in the next time period.  The dynamic system of a multi-hop network running the M-DWARF algorithm is as follows: 

\begin{alignat}{2}
\Delta_0' &= \Delta_{0} + KF_{1} - KF_0 \nonumber \\
\Delta_1' &= \Delta_{1} + KF_{2} - KF_1 \nonumber \\
\vdots \nonumber \\
\Delta_{n-1}' &= \Delta_{n-1} + KF_{0} - KF_{n-1}
\end{alignat}

We write the transition of $\Delta_i$ in a general form as the following,
\begin{alignat}{2}
\Delta_i' = \Delta_{i} + KF_{[i+1]_n} - KF_i. 
\label{eq:transition}
\end{alignat}

Then, we linearly approximate the dynamic system at the equilibrium.

\subsubsection{Linear Approximation}
The Jacobian ($J$) of a difference equations system is defined as follows:

\begin{alignat}{2}
J=\begin{pmatrix} 
\frac{\partial \Delta_{0}'}{\partial \Delta_{0}}  & \frac{\partial \Delta_{0}'}{\partial \Delta_{1}}  & \cdots & \frac{\partial \Delta_{0'}}{\partial \Delta_{n-1}} \\ 
\frac{\partial \Delta_{1}'}{\partial \Delta_{0}}  & \frac{\partial \Delta_{1}'}{\partial \Delta_{1}}   & \cdots & \frac{\partial \Delta_{1}'}{\partial \Delta_{n-1}} \\ 
\vdots & \vdots & \ddots & \vdots \\
\frac{\partial \Delta_{n-1}'}{\partial \Delta_{0}}  & \frac{\partial \Delta_{n-1}'}{\partial \Delta_{1}} & \cdots &  \frac{\partial \Delta_{n-1}'}{\partial \Delta_{n-1}}
\end{pmatrix}
\end{alignat}

Therefore, from Equation \ref{eq:transition}, each element in a row of Jacobian is
\begin{alignat}{2}
\frac{\partial \Delta_i'}{\partial \Delta_p} = \frac{\partial\Delta_{i}}{\partial \Delta_p} + K\frac{\partial F_{[i+1]_n}}{\partial \Delta_p} - K\frac{\partial F_i}{\partial \Delta_p}, p \in \{0,1,\cdots,n-1\}.
\label{eq:element}
\end{alignat}

For the first term, $\frac{\partial\Delta_{i}}{\partial \Delta_p}$ is 1 if $p = i$. . If $p \neq i$, this term is zero. Formally,
\begin{alignat}{2}
\frac{\partial\Delta_{i}}{\partial \Delta_p} = \begin{cases}1 & \text{if }p = i \\ 0 & \text{otherwise.}\end{cases}
\label{eq:diffdelta}
\end{alignat}

Then, we find $\frac{\partial F_i}{\partial \Delta_p}$ as follows:

\begin{alignat}{2}
\frac{\partial F_i}{\partial \Delta_i} =& \text{ }T \Bigg( \sum_{j= i + 2}^{i+(\frac{n}{2}-2)}  \frac{c_{i,[j]_n}(R_{i,j} + S_{i,j} - T_{i,j})}{(\sum_{k=i}^{j-1}\Delta_{[k]_n})^2} \nonumber \\
&+ \frac{c_{i,[i+(\frac{n}{2}-1)]_n}(R_{i,i+(\frac{n}{2}-1)} - T_{i,i+(\frac{n}{2}-1)})}{(\sum_{k=i}^{i+(\frac{n}{2}-2)}\Delta_{[k]_n})^2} \nonumber \\
&+ \frac{c_{i,[i+1]_n}(R_{i,i+1} + S_{i,i+1})}{(\Delta_i)^2} \Bigg) \nonumber \\
\frac{\partial F_i}{\partial \Delta_{[m]_n}} =& \text{ }T \Bigg( \sum_{j= m + 1}^{i+(\frac{n}{2}-2)}  \frac{c_{i,[j]_n}(R_{i,j} + S_{i,j} - T_{i,j})}{(\sum_{k=i}^{j-1}\Delta_{[k]_n})^2} \nonumber \\
&+ \frac{c_{i,[i+(\frac{n}{2}-1)]_n}(R_{i,i+(\frac{n}{2}-1)} - T_{i,i+(\frac{n}{2}-1)})}{(\sum_{k=i}^{i+(\frac{n}{2}-2)}\Delta_{[k]_n})^2} \Bigg), \nonumber \\
&\text{for } m = i+1, i+2, \cdots, i + (\frac{n}{2} - 3) \nonumber  \\
\frac{\partial F_i}{\partial \Delta_{[i+(\frac{n}{2}-2)]_n}} =& \text{ } T \Bigg( \frac{c_{i,[i+(\frac{n}{2}-1)]_n}(R_{i,i+(\frac{n}{2}-1)} - T_{i,i+(\frac{n}{2}-1)})}{(\sum_{k=i}^{i+(\frac{n}{2}-2)}\Delta_{[k]_n})^2} \Bigg) \nonumber \ \\
\frac{\partial F_i}{\partial \Delta_{[m]_n}} =& \text{ }0, \text{ for } m = i + (\frac{n}{2} - 1), i + \frac{n}{2} \nonumber \\
\frac{\partial F_i}{\partial \Delta_{[i+(\frac{n}{2}+1)]_n}} =& \text{ } T \Bigg( - \frac{c_{i,[i+(\frac{n}{2}+1)]_n}(R_{i,i+(\frac{n}{2}+1)} - T_{i,i+(\frac{n}{2}+1)})}{(\sum_{k=i+(\frac{n}{2}+1)}^{i+(n-1)}\Delta_{[k]_n})^2} \Bigg) \nonumber \\
\frac{\partial F_i}{\partial \Delta_{[m]_n}} =& \text{ }T \Bigg(- \sum_{j= i + (\frac{n}{2}+2)}^{m}  \frac{c_{i,[j]_n}(R_{i,j} + S_{i,j} - T_{i,j})}{(\sum_{k=j}^{i-1}\Delta_{[k]_n})^2} \nonumber \\
&- \frac{c_{i,[i+(\frac{n}{2}+1)]_n}(R_{i,i+(\frac{n}{2}+1)} - T_{i,i+(\frac{n}{2}+1)})}{(\sum_{k=i+(\frac{n}{2}+1)}^{i+(n-1)}\Delta_{[k]_n})^2} \Bigg), \nonumber \displaybreak\\
&\text{for } m = i+(\frac{n}{2}+2), i+(\frac{n}{2}+3), \cdots, i + (n-2) \nonumber \\
\frac{\partial F_i}{\partial \Delta_{[i+(n-1)]_n}} =& \text{ }T \Bigg(- \sum_{j= i + (\frac{n}{2}+ 2)}^{i+(n-2)}  \frac{c_{i,[j]_n}(R_{i,j} + S_{i,j} - T_{i,j})}{(\sum_{k=j}^{i-1}\Delta_{[k]_n})^2} \nonumber \\
&- \frac{c_{i,[i+(\frac{n}{2}+1)]_n}(R_{i,i+(\frac{n}{2}+1)} - T_{i,i+(\frac{n}{2}+1)})}{(\sum_{k=i+(\frac{n}{2}+1)}^{i+(n-1)}\Delta_{[k]_n})^2} \nonumber \\
&+ \frac{c_{i,[i+(n-1)]_n}(R_{i,i+(n-1)} + S_{i,i+(n-1)})}{(\Delta_{[i+(n-1)]_n})^2} \Bigg).
\label{eq:difffi}
\end{alignat}

For $\frac{\partial F_{[i+1]_n}}{\partial \Delta_p}$, we can find by replacing $i$ with $i+1$ in Equation \ref{eq:difffi}. The result is as follows:

\begin{alignat}{2}
\frac{\partial F_{[i+1]_n}}{\partial\Delta_{[i+1]_n}} =& \text{ }T \Bigg( \sum_{j= i+3}^{i+(\frac{n}{2}-1)}  \frac{c_{[i+1]_n,[j]_n}(R_{i+1,j} + S_{i+1,j} - T_{i+1,j})}{(\sum_{k=i+1}^{j-1}\Delta_{[k]_n})^2} \nonumber \\
&+ \frac{c_{[i+1]_n,[i+\frac{n}{2}]_n}(R_{i+1,i+\frac{n}{2}} - T_{i+1,i+\frac{n}{2}})}{(\sum_{k=i+1}^{i+(\frac{n}{2}-1)}\Delta_{[k]_n})^2} \nonumber \\
&+ \frac{c_{[i+1]_n,[i+2]_n}(R_{i+1,i+2} + S_{i+1,i+2})}{(\Delta_{[i+1]_n})^2} \Bigg) \nonumber \\
\frac{\partial F_{[i+1]_n}}{\partial\Delta_{[m]_n}} =& \text{ }T \Bigg( \sum_{j= m + 1}^{i+(\frac{n}{2}-1)}  \frac{c_{[i+1]_n,[j]_n}(R_{i+1,j} + S_{i+1,j} - T_{i+1,j})}{(\sum_{k=i+1}^{j-1}\Delta_{[k]_n})^2} \nonumber \\
&+ \frac{c_{[i+1]_n,[i+\frac{n}{2}]_n}(R_{i+1,i+\frac{n}{2}} - T_{i+1,i+\frac{n}{2}})}{(\sum_{k=i+1}^{i+(\frac{n}{2}-1)}\Delta_{[k]_n})^2} \Bigg), \nonumber \\
&\text{for } m = i+2, i+3, \cdots, i+ (\frac{n}{2} - 2) \nonumber  \\
\frac{\partial F_{[i+1]_n}}{\partial\Delta_{i+(\frac{n}{2}-1)}} =& \text{ } T \Bigg( \frac{c_{[i+1]_n,[i+\frac{n}{2}]_n}(R_{i+1,i+\frac{n}{2}} - T_{i+1,i+\frac{n}{2}})}{(\sum_{k=i+1}^{i+(\frac{n}{2}-1)}\Delta_{[k]_n})^2} \Bigg) \nonumber \ \\
\frac{\partial F_{[i+1]_n}}{\partial\Delta_[m]_n} =& \text{ }0, \text{ for } m = i+ \frac{n}{2}, i+( \frac{n}{2} + 1) \nonumber \\
\frac{\partial F_{[i+1]_n}}{\partial\Delta_{[i+(\frac{n}{2}+2)]_n}} =& \text{ } T \Bigg( - \frac{c_{[i+1]_n,[i+(\frac{n}{2}+2)]_n}(R_{i+1,i+(\frac{n}{2}+2)} - T_{i+1,i+(\frac{n}{2}+2)})}{(\sum_{k=i+(\frac{n}{2}+2)}^{i+n}\Delta_{[k]_n})^2} \Bigg) \nonumber \\
\frac{\partial F_{[i+1]_n}}{\partial\Delta_{[m]_n}} =& \text{ }T \Bigg(- \sum_{j= i+ (\frac{n}{2}+3)}^{m}  \frac{c_{[i+1]_n,[j]_n}(R_{i+1,j} + S_{i+1,j} - T_{i+1,j})}{(\sum_{k=j}^{i}\Delta_{[k]_n})^2} \nonumber \\
&- \frac{c_{[i+1]_n,[i+(\frac{n}{2}+2)]_n}(R_{i+1,i+(\frac{n}{2}+2)} - T_{i+1,i+(\frac{n}{2}+2)})}{(\sum_{k=i+(\frac{n}{2}+2)}^{i+n}\Delta_{[k]_n})^2} \Bigg), \nonumber \\
&\text{for } m = i+(\frac{n}{2}+3), i+(\frac{n}{2}+4), \cdots, i+ (n-1) \nonumber \\
\frac{\partial F_{[i+1]_n}}{\partial\Delta_{i}} =& \text{ }T \Bigg(- \sum_{j= i + (\frac{n}{2}+ 3)}^{i+(n-1)}  \frac{c_{[i+1]_n,[j]_n}(R_{i+1,j} + S_{i+1,j} - T_{i+1,j})}{(\sum_{k=j}^{i}\Delta_{[k]_n})^2} \nonumber \\
&- \frac{c_{[i+1]_n,[i+(\frac{n}{2}+2)]_n}(R_{i+1,i+(\frac{n}{2}+2)} - T_{i+1,i+(\frac{n}{2}+2)})}{(\sum_{k=i+(\frac{n}{2}+2)}^{i+n}\Delta_{[k]_n})^2} \nonumber \\
&- \frac{c_{[i+1]_n,i}(R_{i+1,i} + S_{i+1,i})}{(\Delta_{i})^2} \Bigg).
\label{eq:difffiplus1}
\end{alignat}
Consequently, from Equation \ref{eq:element}, \ref{eq:diffdelta}, \ref{eq:difffi}, and \ref{eq:difffiplus1}, we find $\frac{\partial \Delta_i'}{\partial \Delta_p}$ as the following:
\begin{alignat}{2}
\frac{\partial \Delta_i'}{\partial \Delta_i} =& \text{ }1 + KT \Bigg(- \sum_{j= i + (\frac{n}{2}+ 3)}^{i+(n-1)}  \frac{c_{[i+1]_n,[j]_n}(R_{i+1,j} + S_{i+1,j} - T_{i+1,j})}{(\sum_{k=j}^{i}\Delta_{[k]_n})^2} \nonumber \\
&- \frac{c_{[i+1]_n,[i+(\frac{n}{2}+2)]_n}(R_{i+1,i+(\frac{n}{2}+2)} - T_{i+1,i+(\frac{n}{2}+2)})}{(\sum_{k=i+(\frac{n}{2}+2)}^{i+n}\Delta_{[k]_n})^2} \nonumber \\
&- \frac{c_{[i+1]_n,i}(R_{i+1,i} + S_{i+1,i})}{(\Delta_{i})^2} \nonumber \\
&- \sum_{j= i + 2}^{i+(\frac{n}{2}-2)}  \frac{c_{i,[j]_n}(R_{i,j} + S_{i,j} - T_{i,j})}{(\sum_{k=i}^{j-1}\Delta_{[k]_n})^2} \nonumber \\
&- \frac{c_{i,[i+(\frac{n}{2}-1)]_n}(R_{i,i+(\frac{n}{2}-1)} - T_{i,i+(\frac{n}{2}-1)})}{(\sum_{k=i}^{i+(\frac{n}{2}-2)}\Delta_{[k]_n})^2} \nonumber \\
&- \frac{c_{i,[i+1]_n}(R_{i,i+1} + S_{i,i+1})}{(\Delta_i)^2} \Bigg) \nonumber   \\
\frac{\partial \Delta_i'}{\partial \Delta_{[i+1]_n}} =& \text{ }KT \Bigg(\sum_{j= i+3}^{i+(\frac{n}{2}-1)}  \frac{c_{[i+1]_n,[j]_n}(R_{i+1,j} + S_{i+1,j} - T_{i+1,j})}{(\sum_{k=i+1}^{j-1}\Delta_{[k]_n})^2} \nonumber \\
&+ \frac{c_{[i+1]_n,[i+\frac{n}{2}]_n}(R_{i+1,i+\frac{n}{2}} - T_{i+1,i+\frac{n}{2}})}{(\sum_{k=i+1}^{i+(\frac{n}{2}-1)}\Delta_{[k]_n})^2} \nonumber \\
&+ \frac{c_{[i+1]_n,[i+2]_n}(R_{i+1,i+2} + S_{i+1,i+2})}{(\Delta_{[i+1]_n})^2} \nonumber \\
&- \sum_{j= i + 2}^{i+(\frac{n}{2}-2)}  \frac{c_{i,[j]_n}(R_{i,j} + S_{i,j} - T_{i,j})}{(\sum_{k=i}^{j-1}\Delta_{[k]_n})^2} \nonumber \\
&- \frac{c_{i,[i+(\frac{n}{2}-1)]_n}(R_{i,[i+(\frac{n}{2}-1)]_n} - T_{i,i+(\frac{n}{2}-1)})}{(\sum_{k=i}^{i+(\frac{n}{2}-2)}\Delta_{[k]_n})^2} \Bigg) \nonumber \\
\frac{\partial \Delta_i'}{\partial \Delta_{[m]_n}} =& \text{ }KT \Bigg(\sum_{j= m+1}^{i+(\frac{n}{2}-1)}  \frac{c_{[i+1]_n,[j]_n}(R_{i+1,j} + S_{i+1,j} - T_{i+1,j})}{(\sum_{k=i+1}^{j-1}\Delta_{[k]_n})^2} \nonumber \\
&+ \frac{c_{[i+1]_n,[i+\frac{n}{2}]_n}(R_{i+1,i+\frac{n}{2}} - T_{i+1,i+\frac{n}{2}})}{(\sum_{k=i+1}^{i+(\frac{n}{2}-1)}\Delta_{[k]_n})^2} \nonumber \\
&- \sum_{j= m+1}^{i+(\frac{n}{2}-2)}  \frac{c_{i,[j]_n}(R_{i,j} + S_{i,j} - T_{i,j})}{(\sum_{k=i}^{j-1}\Delta_{[k]_n})^2} \nonumber \\
&- \frac{c_{i,[i+(\frac{n}{2}-1)]_n}(R_{i,i+(\frac{n}{2}-1)} - T_{i,i+(\frac{n}{2}-1)})}{(\sum_{k=i}^{i+(\frac{n}{2}-2)}\Delta_{[k]_n})^2}, \Bigg), \nonumber \\
&\text{for } m = i+2, i+3, \cdots, i + (\frac{n}{2} - 3) \nonumber   \\
\frac{\partial \Delta_i'}{\partial \Delta_{[i+(\frac{n}{2}-2)]_n}} =& \text{ }KT \Bigg( \frac{c_{[i+1]_n,[i+(\frac{n}{2}-1)]_n}(R_{i+1,i+(\frac{n}{2}-1)} + S_{i+1,i+(\frac{n}{2}-1)} - T_{i+1,i+(\frac{n}{2}-1)})}{(\sum_{k=i+1}^{i+(\frac{n}{2}-2)}\Delta_{[k]_n})^2} \nonumber \\
&+ \frac{c_{[i+1]_n,[i+\frac{n}{2}]_n}(R_{i+1,i+\frac{n}{2}} - T_{i+1,i+\frac{n}{2}})}{(\sum_{k=i+1}^{i+(\frac{n}{2}-1)}\Delta_{[k]_n})^2} \nonumber \\
&- \frac{c_{i,[i+(\frac{n}{2}-1)]_n}(R_{i,i+(\frac{n}{2}-1)} - T_{i,i+(\frac{n}{2}-1)})}{(\sum_{k=i}^{i+(\frac{n}{2}-2)}\Delta_{[k]_n})^2} \Bigg) \nonumber \displaybreak \\
\frac{\partial \Delta_i'}{\partial \Delta_{i+(\frac{n}{2}-1)}} =& \text{ } KT \Bigg( \frac{c_{[i+1]_n,[i+\frac{n}{2}]_n}(R_{i+1,i+\frac{n}{2}} - T_{i+1,i+\frac{n}{2}})}{(\sum_{k=i+1}^{i+(\frac{n}{2}-1)}\Delta_{[k]_n})^2} \Bigg) \nonumber \ \\
\frac{\partial \Delta_i'}{\partial \Delta_{[i+\frac{n}{2}]_n}} =& \text{ }0 \nonumber \\
\frac{\partial \Delta_i'}{\partial \Delta_{[i+(\frac{n}{2}+1)]_n}} =& \text{ } KT \Bigg( \frac{c_{i,[i+(\frac{n}{2}+1)]_n}(R_{i,i+(\frac{n}{2}+1)} - T_{i,i+(\frac{n}{2}+1)})}{(\sum_{k=i+(\frac{n}{2}+1)}^{i+(n-1)}\Delta_{[k]_n})^2} \Bigg) \nonumber  \\
\frac{\partial \Delta_i'}{\partial \Delta_{i+(\frac{n}{2}+2)}} =& \text{ }KT \Bigg( - \frac{c_{[i+1]_n,[i+(\frac{n}{2}+2)]_n}(R_{i+1,i+(\frac{n}{2}+2)} - T_{i+1,i+(\frac{n}{2}+2)})}{(\sum_{k=i+(\frac{n}{2}+2)}^{i+n}\Delta_{[k]_n})^2} \nonumber \\
&+ \frac{c_{i,[i+(\frac{n}{2}+2)]_n}(R_{i,i+(\frac{n}{2}+2)} + S_{i,i+(\frac{n}{2}+2)} - T_{i,i+(\frac{n}{2}+2)})}{(\sum_{k=i+(\frac{n}{2}+2)}^{i+(n-1)}\Delta_{[k]_n})^2} \nonumber \\
&+ \frac{c_{i,[i+(\frac{n}{2}+1)]_n}(R_{i,i+(\frac{n}{2}+1)} - T_{i,i+(\frac{n}{2}+1)})}{(\sum_{k=i+(\frac{n}{2}+1)}^{i+(n-1)}\Delta_{[k]_n})^2} \Bigg) \nonumber \\
\frac{\partial \Delta_i'}{\partial \Delta_{[m]_n}} =& \text{ }KT \Bigg(- \sum_{j= i + (\frac{n}{2}+3)}^{m}  \frac{c_{[i+1]_n,[j]_n}(R_{i+1,j} + S_{i+1,j} - T_{i+1,j})}{(\sum_{k=j}^{i}\Delta_{[k]_n})^2} \nonumber \\
&- \frac{c_{[i+1]_n,[i+(\frac{n}{2}+2)]_n}(R_{i+1,i+(\frac{n}{2}+2)} - T_{i+1,i+(\frac{n}{2}+2)})}{(\sum_{k=i+(\frac{n}{2}+2)}^{i+n}\Delta_{[k]_n})^2} \nonumber \\
&+\sum_{j= i + (\frac{n}{2}+2)}^{m}  \frac{c_{i,[j]_n}(R_{i,j} + S_{i,j} - T_{i,j})}{(\sum_{k=j}^{i+(n-1)}\Delta_{[k]_n})^2} \nonumber \\
&+ \frac{c_{i,[i+(\frac{n}{2}+1)]_n}(R_{i,i+(\frac{n}{2}+1)} - T_{i,i+(\frac{n}{2}+1)})}{(\sum_{k=i+(\frac{n}{2}+1)}^{i+(n-1)}\Delta_{[k]_n})^2}\Bigg), \nonumber \\
&\text{for } m = i+(\frac{n}{2}+3), i+(\frac{n}{2}+4), \cdots, i + (n-2) \nonumber \\
\frac{\partial \Delta_i'}{\partial \Delta_{[i+(n-1)]_n}} =& \text{ }KT \Bigg(- \sum_{j= i + (\frac{n}{2}+ 3)}^{i+(n-1)}  \frac{c_{[i+1]_n,[j]_n}(R_{i+1,j} + S_{i+1,j} - T_{i+1,j})}{(\sum_{k=j}^{i}\Delta_{[k]_n})^2} \nonumber \\
&- \frac{c_{[i+1]_n,[i+(\frac{n}{2}+2)]_n}(R_{i+1,i+(\frac{n}{2}+2)} - T_{i+1,i+(\frac{n}{2}+2)})}{(\sum_{k=i+(\frac{n}{2}+2)}^{i+n}\Delta_{[k]_n})^2} \nonumber \\
&+\sum_{j= i + (\frac{n}{2}+2)}^{i+(n-2)}  \frac{c_{i,[j]_n}(R_{i,j} + S_{i,j} - T_{i,j})}{(\sum_{k=j}^{i+(n-1)}\Delta_{[k]_n})^2} \nonumber \\
&+ \frac{c_{i,[i+(\frac{n}{2}+1)]_n}(R_{i,i+(\frac{n}{2}+1)} - T_{i,i+(\frac{n}{2}+1)})}{(\sum_{k=i+(\frac{n}{2}+1)}^{i+(n-1)}\Delta_{[k]_n})^2} \nonumber \\
&+ \frac{c_{i,[i+(n-1)]_n}(R_{i,i+(n-1)} + S_{i,i+(n-1)})}{(\Delta_{[i+(n-1)]_n})^2} \Bigg).
\label{eq:jacobiandiff}
\end{alignat}

At this point, we already get the general form of each row of the Jacobian matrix. To locally approximate at the equilibrium, we have to substitute $c_{i,j}, R_{i,j}, S_{i,j}, T_{i,j}$, and $\Delta_i$ for all $i,j \in {0, 1, \cdots, n-1}$ where $j \neq i$ with the value at the equilibrium state. Then, we have find the maximum value of $n$ that bounds the eigenvalues within a unit circle to guarantee the stability. However, such values depend on the topology. For example, the connectivity $c_{i,j}$ depends on whether node $i$ can perceive the presence of node $j$ or not. If node $i$ and $j$ are within two-hop communication, this value is 1, otherwise, this value is 0.

In this paper, we prove the stability of the multi-hop star topology as an example because this topology is less complicated to prove and may be the easiest example for the reader to follow the proof. For other topologies, we conjecture that the local stability can also be assured as well.

For the multi-hop star topology, the time interval between arbitrary two consecutive nodes is $T/n$ at the equilibrium. Therefore, each row of the Jacobian matrix from Equation \ref{eq:jacobiandiff} is changed by substituting all $\Delta_{k}$ with $T/n$ and re-indexing the summation as follows,

\begin{alignat}{2}
\frac{\partial \Delta_i'}{\partial \Delta_i} =& \text{ }1 + \frac{Kn^2}{T} \Bigg(
- \frac{c_{[i+1]_n,[i+(\frac{n}{2}+2)]_n}(R_{i+1,i+(\frac{n}{2}+2)} - T_{i+1,i+(\frac{n}{2}+2)})}{(\frac{n}{2}-1)^2} \nonumber \\
&- \frac{c_{[i+1]_n,[i]_n}(R_{i+1,i} + S_{i+1,i})}{1^2} \nonumber \\
&- \sum_{j=2}^{\frac{n}{2}-2}  \frac{c_{[i+1]_n,[i+(n+1-j)]_n}(R_{i+1,i+(n+1-j)} + S_{i+1,i+(n+1-j)} - T_{i+1,i+(n+1-j)})}{j^2} \nonumber \\
&- \sum_{j= 2}^{\frac{n}{2}-2}  \frac{c_{i,[i+j]_n}(R_{i,i+j} + S_{i,i+j} - T_{i,i+j})}{j^2} \nonumber \\
&- \frac{c_{i,[i+(\frac{n}{2}-1)]_n}(R_{i,i+(\frac{n}{2}-1)} - T_{i,i+(\frac{n}{2}-1)})}{(\frac{n}{2}-1)^2} \nonumber \\
&- \frac{c_{i,[i+1]_n}(R_{i,i+1} + S_{i,i+1})}{1^2} \Bigg)\nonumber \\
\frac{\partial \Delta_i'}{\partial \Delta_{[i+1]_n}} =& \text{ }\frac{Kn^2}{T} \Bigg(\sum_{j=2}^{\frac{n}{2}-2}  \frac{c_{[i+1]_n,[i+1+j]_n}(R_{i+1,i+1+j} + S_{i+1,i+1+j} - T_{i+1,i+1+j})}{j^2} \nonumber \\
&+ \frac{c_{[i+1]_n,[i+\frac{n}{2}]_n}(R_{i+1,i+\frac{n}{2}} - T_{i+1,i+\frac{n}{2}})}{(\frac{n}{2}-1)^2} \nonumber \\
&+ \frac{c_{[i+1]_n,[i+2]_n}(R_{i+1,i+2} + S_{i+1,i+2})}{1^2} \nonumber \\
&- \sum_{j= 2}^{\frac{n}{2}-2}  \frac{c_{i,[i+j]_n}(R_{i,i+j} + S_{i,i+j} - T_{i,i+j})}{j^2} \nonumber \\
&- \frac{c_{i,[i+(\frac{n}{2}-1)]_n}(R_{i,i+(\frac{n}{2}-1)} - T_{i,i+(\frac{n}{2}-1)})}{(\frac{n}{2}-1)^2} \Bigg) \nonumber  \\
\frac{\partial \Delta_i'}{\partial \Delta_{[m]_n}} =& \text{ }\frac{Kn^2}{T} \Bigg(\sum_{j= m-i}^{\frac{n}{2}-2}  \frac{c_{[i+1]_n,[i+1+j]_n}(R_{i+1,i+1+j} + S_{i+1,i+1+j} - T_{i+1,i+1+j})}{j^2} \nonumber \\
&+ \frac{c_{[i+1]_n,[i+\frac{n}{2}]_n}(R_{i+1,i+\frac{n}{2}} - T_{i+1,i+\frac{n}{2}})}{(\frac{n}{2}-1)^2} \nonumber \\
&- \sum_{j= m-i+1}^{\frac{n}{2}-2}  \frac{c_{i,[i+j]_n}(R_{i,i+j} + S_{i,i+j} - T_{i,i+j})}{j^2} \nonumber \\
&- \frac{c_{i,[i+(\frac{n}{2}-1)]_n}(R_{i,i+(\frac{n}{2}-1)} - T_{i,i+(\frac{n}{2}-1)})}{(\frac{n}{2}-1)^2} \Bigg) \nonumber \\
&\text{for } m = i+2, i+3, \cdots, i + (\frac{n}{2} - 3) \nonumber  \\
\frac{\partial \Delta_i'}{\partial \Delta_{[i+(\frac{n}{2}-2)]_n}} =& \text{ }\frac{Kn^2}{T} \Bigg( \frac{c_{[i+1]_n,[i+(\frac{n}{2}-1)]_n}(R_{i+1,i+(\frac{n}{2}-1)} + S_{i+1,i+(\frac{n}{2}-1)} - T_{i+1,i+(\frac{n}{2}-1)})}{(\frac{n}{2}-2)^2} \nonumber \\
&+ \frac{c_{[i+1]_n,[i+\frac{n}{2}]_n}(R_{i+1,i+\frac{n}{2}} - T_{i+1,i+\frac{n}{2}})}{(\frac{n}{2}-1)^2} \nonumber \\
&- \frac{c_{i,[i+(\frac{n}{2}-1)]_n}(R_{i,i+(\frac{n}{2}-1)} - T_{i,i+(\frac{n}{2}-1)})}{(\frac{n}{2}-1)^2} \Bigg) \nonumber \displaybreak \\
\frac{\partial \Delta_i'}{\partial \Delta_{[i+(\frac{n}{2}-1)]_n}} =& \text{ } \frac{Kn^2}{T} \Bigg( \frac{c_{[i+1]_n,[i+\frac{n}{2}]_n}(R_{i+1,i+\frac{n}{2}} - T_{i+1,i+\frac{n}{2}})}{(\frac{n}{2}-1)^2} \Bigg) \nonumber \ \\
\frac{\partial \Delta_i'}{\partial \Delta_{[i+\frac{n}{2}]_n}} =& \text{ }0 \nonumber \\
\frac{\partial \Delta_i'}{\partial \Delta_{[i+(\frac{n}{2}+1)]_n}} =& \text{ } \frac{Kn^2}{T} \Bigg( \frac{c_{i,[i+(\frac{n}{2}+1)]_n}(R_{i,i+(\frac{n}{2}+1)} - T_{i,i+(\frac{n}{2}+1)})}{(\frac{n}{2}-1)^2} \Bigg) \nonumber \\
\frac{\partial \Delta_i'}{\partial \Delta_{[i+(\frac{n}{2}+2)]_n}} =& \text{ }\frac{Kn^2}{T} \Bigg( - \frac{c_{[i+1]_n,[i+(\frac{n}{2}+2)]_n}(R_{i+1,i+(\frac{n}{2}+2)} - T_{i+1,i+(\frac{n}{2}+2)})}{(\frac{n}{2}-1)^2} \nonumber \\
&+ \frac{c_{i,[i+(\frac{n}{2}+2)]_n}(R_{i,i+(\frac{n}{2}+2)} + S_{i,i+(\frac{n}{2}+2)} - T_{i,i+(\frac{n}{2}+2)})}{(\frac{n}{2}-2)^2} \nonumber \\
&+ \frac{c_{i,[i+(\frac{n}{2}+1)]_n}(R_{i,i+(\frac{n}{2}+1)} - T_{i,i+(\frac{n}{2}+1)})}{(\frac{n}{2}-1)^2} \Bigg) \nonumber \\
\frac{\partial \Delta_i'}{\partial \Delta_{[m]_n}} =& \text{ }\frac{Kn^2}{T} \Bigg(- \frac{c_{[i+1]_n,[i+(\frac{n}{2}+2)]_n}(R_{i+1,i+(\frac{n}{2}+2)} - T_{i+1,i+(\frac{n}{2}+2)})}{(\frac{n}{2}-1)^2} \nonumber \\
&- \sum_{j= i + n -m + 1}^{\frac{n}{2}-2}  \frac{c_{[i+1]_n,[j-\frac{n}{2}+2+m]_n}(R_{i+1,j-\frac{n}{2}+2+m} + S_{i+1,j-\frac{n}{2}+2+m} - T_{i+1,j-\frac{n}{2}+2+m})}{j^2} \nonumber \\
&+ \sum_{j= i + n - m}^{\frac{n}{2}-2}  \frac{c_{i,[j-\frac{n}{2}+2+m]_n}(R_{i,j-\frac{n}{2}+2+m} + S_{i,j-\frac{n}{2}+2+m} - T_{i,j-\frac{n}{2}+2+m})}{j^2} \nonumber \\
&+ \frac{c_{i,[i+(\frac{n}{2}+1)]_n}(R_{i,i+(\frac{n}{2}+1)} - T_{i,i+(\frac{n}{2}+1)})}{(\frac{n}{2}-1)^2}\Bigg), \nonumber \\
&\text{for } m = i+(\frac{n}{2}+3), i+(\frac{n}{2}+4), \cdots, i + (n-2) \nonumber \\
\frac{\partial \Delta_i'}{\partial \Delta_{[i+(n-1)]_n}} =& \text{ }\frac{Kn^2}{T} \Bigg(- \frac{c_{[i+1]_n,[i+(\frac{n}{2}+2)]_n}(R_{i+1,i+(\frac{n}{2}+2)} - T_{i+1,i+(\frac{n}{2}+2)})}{(\frac{n}{2}-1)^2} \nonumber \\
&- \sum_{j= 2}^{\frac{n}{2}-2}  \frac{c_{[i+1]_n,[i+\frac{n}{2}+1+j]_n}(R_{i+1,i+\frac{n}{2}+1+j} + S_{i+1,i+\frac{n}{2}+1+j} - T_{i+1,i+\frac{n}{2}+1+j})}{j^2} \nonumber \\
&+ \sum_{j=2}^{\frac{n}{2}-2}  \frac{c_{i,[i+\frac{n}{2}+j]_n}(R_{i,i+\frac{n}{2}+j} + S_{i,i+\frac{n}{2}+j} - T_{i,i+\frac{n}{2}+j})}{j^2} \nonumber \\
&+ \frac{c_{i,[i+(\frac{n}{2}+1)]_n}(R_{i,i+(\frac{n}{2}+1)} - T_{i,i+(\frac{n}{2}+1)})}{(\frac{n}{2}-1)^2} \nonumber \\
&+ \frac{c_{i,[i+(n-1)]_n}(R_{i,i+(n-1)} + S_{i,i+(n-1)})}{1^2} \Bigg) \nonumber.
\label{eq:startndiff}
\end{alignat}

Then, for the star topology, nodes perceive the presence of all two-hop neighbors. Therefore, all terms $c_{i,j}, R_{i,j}, S_{i,j}$, and $T_{i,j}$ are 1. Consequently, the row of the Jacobian matrix at the equilibrium is the following: 

\begin{alignat}{2}
\frac{\partial \Delta_i'}{\partial \Delta_i} =& \text{ }1 + \frac{Kn^2}{T} \Bigg(- \sum_{j=2}^{\frac{n}{2}-2}  \frac{1}{j^2} - 0 - \frac{2}{1^2} - \sum_{j= 2}^{\frac{n}{2}-2}  \frac{1}{j^2} - 0 - \frac{2}{1^2} \Bigg) \nonumber \\
&= 1 - \frac{2Kn^2}{T}\Bigg(1 + \sum_{j=1}^{n-2}\frac{1}{j^2}\Bigg) \nonumber \displaybreak \\
\frac{\partial \Delta_i'}{\partial \Delta_{[i+1]_n}} =& \text{ }\frac{Kn^2}{T} \Bigg(\sum_{j=2}^{\frac{n}{2}-2}  \frac{1}{j^2} + 0 + 2 - \sum_{j= 2}^{\frac{n}{2}-2}  \frac{1}{j^2}- 0 \Bigg) = \frac{2Kn^2}{T} \nonumber \\
\frac{\partial \Delta_i'}{\partial \Delta_{[m]_n}} =& \text{ }\frac{Kn^2}{T} \Bigg(\sum_{j= m-i}^{\frac{n}{2}-2}  \frac{1}{j^2} + 0 - \sum_{j= m-i+1}^{\frac{n}{2}-2}  \frac{1}{j^2} - 0 \Bigg) = \frac{Kn^2}{T(m-i)^2} \nonumber \\
&\text{for } m = i+2, i+3, \cdots, i + (\frac{n}{2} - 3) \nonumber  \\
\frac{\partial \Delta_i'}{\partial \Delta_{[i+(\frac{n}{2}-2)]_n}} =& \text{ }\frac{Kn^2}{T} \Bigg( \frac{1}{(\frac{n}{2}-2)^2} + 0 - 0 \Bigg) = \frac{Kn^2}{T(\frac{n}{2}-2)^2} \nonumber \\
\frac{\partial \Delta_i'}{\partial \Delta_{[i+(\frac{n}{2}-1)]_n}} =& \text{ }0 \nonumber \ \\
\frac{\partial \Delta_i'}{\partial \Delta_{[i+\frac{n}{2}]_n}} =& \text{ }0 \nonumber \\
\frac{\partial \Delta_i'}{\partial \Delta_{[i+(\frac{n}{2}+1)]_n}} =& \text{ } 0 \nonumber \\
\frac{\partial \Delta_i'}{\partial \Delta_{[i+(\frac{n}{2}+2)]_n}} =& \text{ }\frac{Kn^2}{T} \Bigg( - 0 + \frac{1}{(\frac{n}{2}-2)^2} + 0 \Bigg) =  \frac{Kn^2}{T(\frac{n}{2}-2)^2} \nonumber \\
\frac{\partial \Delta_i'}{\partial \Delta_{[m]_n}} =& \text{ }\frac{Kn^2}{T} \Bigg(- \sum_{j= i + n -m + 1}^{\frac{n}{2}-2}  \frac{1}{j^2} - 0 + \sum_{j= i + n - m}^{\frac{n}{2}-2}  \frac{1}{j^2} + 0 = \frac{Kn^2}{T(i+n-m)^2}, \nonumber \\
&\text{ for } m = i+(\frac{n}{2}+3), i+(\frac{n}{2}+4), \cdots, i + (n-2) \nonumber \\
\frac{\partial \Delta_i'}{\partial \Delta_{[i+(n-1)]_n}} =& \text{ }\frac{Kn^2}{T} \Bigg(- \sum_{j= 2}^{\frac{n}{2}-2}  \frac{1}{j^2} - 0 + \sum_{j=2}^{\frac{n}{2}-2}  \frac{1}{j^2} + 0 + \frac{2}{1^2} \Bigg) = \frac{2Kn^2}{T}. 
\end{alignat}

The result of the Jacobian matrix at the equilibrium is the circulant matrix as shown below:

\setcounter{MaxMatrixCols}{11}
\begin{alignat}{2}
\begin{pmatrix} 
D_0 & 2A  & \cdots & \frac{A}{(\frac{n}{2}-2)^2} & 0 & 0 & 0 & \frac{A}{(\frac{n}{2}-2)^2} & \cdots  & \frac{A}{2^2} & 2A \\
2A & D_0 & \cdots & \frac{A}{(\frac{n}{2}-3)^2} & \frac{A}{(\frac{n}{2}-2)^2} & 0 & 0 & 0  & \cdots & \frac{A}{3^2} & \frac{A}{2^2} \\
\vdots & \vdots &  \ddots &  \vdots & \vdots & \vdots & \vdots & \vdots & \ddots & \vdots & \vdots \\
\frac{A}{(\frac{n}{2}-2)^2} & \frac{A}{(\frac{n}{2}-3)^2} & \cdots & D_0 & 2A & \frac{A}{2^2} & \frac{A}{3^2} & \frac{A}{4^2} & \cdots & 0 & 0  \\
0 & \frac{A}{(\frac{n}{2}-2)^2} & \cdots  & 2A & D_0 & 2A & \frac{A}{2^2} & \frac{A}{3^2} & \cdots & 0 & 0 \\
0 & 0 & \cdots  & \frac{A}{2^2} & 2A & D_0 & 2A & \frac{A}{2^2} & \cdots & \frac{A}{(\frac{n}{2}-2)^2} & 0  \\
0 & 0  & \cdots  & \frac{A}{3^2} & \frac{A}{2^2} & 2A & D_0 & 2A &  \cdots & \frac{A}{(\frac{n}{2}-3)^2} & \frac{A}{(\frac{n}{2}-2)^2}  \\
\frac{A}{(\frac{n}{2}-2)^2} & 0  & \cdots  & \frac{A}{4^2} & \frac{A}{3^2} & \frac{A}{2^2} & 2A &  D_0 &  \cdots & \frac{A}{(\frac{n}{2}-4)^2} & \frac{A}{(\frac{n}{2}-3)^2}  \\
\vdots & \vdots & \ddots &  \vdots & \vdots & \vdots &  \vdots & \vdots &  \ddots  &  \vdots & \vdots\\
\frac{A}{2^2} &  \frac{A}{3^2} &\cdots & 0 & 0 & \frac{A}{(\frac{n}{2}-2)^2} & \frac{A}{(\frac{n}{2}-3)^2} & \frac{A}{(\frac{n}{2}-3)^2} & \cdots & D_0 & 2A\\
2A & \frac{A}{2^2}  & \cdots & 0 & 0 & 0 & \frac{A}{(\frac{n}{2}-2)^2}& \frac{A}{(\frac{n}{2}-3)^2} & \cdots  & 2A & D_0
\end{pmatrix},
\end{alignat}

where the diagonal entry $D_0 = 1 - 2A\Bigg(1 + \sum_{j=1}^{n-2}1/j^2\Bigg)$ and $A=Kn^2/T$.

We can find each eigenvalue of a circulant matrix with a general solution presented in \cite{circulant}. However, we only need to guarantee that all eigenvalues lay on a unit circle. Therefore, we find the bound of eigenvalues instead.

\subsubsection{The Bound of Eigenvalues}
\label{sec:bound-mhop}
To find the bound of an $n \times n$ matrix, we use the Gershgorin's Circle Theorem (\cite{gersheng,gershger})

\begin{theorem}[Gershgorin’s Theorem Round 1]
	Every eigenvalue $\lambda$ of $n \times n$ matrix $A$ satisfies:
	
	\begin{alignat}{2}
	|\lambda - A_{i,i}| \leq \sum_{j \neq i} |A_{i,j}|, \text{ } i \in {0,1,\cdots, n-1} \nonumber
	\end{alignat}
\end{theorem}

In other words, every eigenvalue lies within at least one of Gershgorin discs centered at $A_{i,i}$ with radius $\sum_{j \neq i} |A_{i,j}|$, where $A_{i,i}$ is the diagonal entry of a matrix.

In our circulant matrix, all diagonal entries and the sums of elements in each row and each column are the same. Therefore, in our matrix, all Gershgorin's discs are centered at $D_0 = 1 - 2A(1 + \sum_{j=1}^{n-2}1/j^2)$ with radius $r = 2A(1 + \sum_{j=1}^{\frac{n}{2}-2}1/j^2)$, where $A = Kn^2/T$.

Then, we find the maximum number of nodes $n$ that guarantees the Gershgorin' discs are in a unit circle.

Let $\vec{D}$ be a vector drawn from the origin $(0,0)$ to the center of the Gershgorin's disc $(1 - 2A(1 + \sum_{j=1}^{n-2}1/j^2), 0)$. Due to the imaginary part of $\vec{D}$ is zero, the magnitude $|\vec{D}|$ is $|1 - 2A(1 + \sum_{j=1}^{n-2}1/j^2|$. Therefore, we derive the following to guarantee the Gershgorin's discs are in a unit circle:

\begin{alignat}{2}
|\vec{D}| + r \leq& \text{ } 1 \nonumber \\
\Bigg|1 - 2A\Bigg(1 + \sum_{j=1}^{n-2}\frac{1}{j^2}\Bigg)\Bigg| + 2A\Bigg(1 + \sum_{j=1}^{\frac{n}{2}-2}\frac{1}{j^2}\Bigg) \leq& \text{ } 1 \nonumber \\
\Bigg|1 - \frac{2Kn^2}{T}\Bigg(1 + \sum_{j=1}^{n-2}\frac{1}{j^2}\Bigg)\Bigg| + \frac{2Kn^2}{T}\Bigg(1 + \sum_{j=1}^{\frac{n}{2}-2}\frac{1}{j^2}\Bigg) \leq& \text{ } 1.
\label{eq:disc}
\end{alignat}

From the M-DWARF algorithm, we substitute $K$ with $38.597 \times n^{-1.874} \times T/1000$. Additionally, when $n$ is large, the value of $\sum_{j=1}^{n-2} 1/j^2$ and $\sum_{j=1}^{\frac{n}{2}-2}1/j^2$ converge to the Reimann zeta function $\zeta (2) = \sum_{i=1}^{\infty}1/i^2 = \pi^2/6 \approx 1.645$. From Equation \ref{eq:disc}, we get the following:

\begin{alignat}{2}
\Bigg|1 - 0.077194n^{0.126}(1+ 1.645)\Bigg| +0.077194n^{0.126}(1+1.645) &\leq \text{ } 1 \nonumber \\
\Bigg|1 - 0.20417813n^{0.126}\Bigg| + 0.20417813n^{0.126} &\leq \text{ } 1 \nonumber \\
- (1 - 0.20417813n^{0.126}) \leq 1 - 0.20417813n^{0.126}  &\leq \text{ } 1 - 0.20417813n^{0.126}\nonumber \\
\end{alignat}

The condition $1 - 0.20417813n^{0.126}  \leq \text{ } 1 - 0.20417813n^{0.126}$ is always true regardless of the number of nodes $n$. Then, we consider the following condition:
\begin{alignat}{2}
- (1 - 0.20417813n^{0.126}) &\leq 1 - 0.20417813n^{0.126} \nonumber \\
2(0.20417813n^{0.126} - 1) &\leq 0 \nonumber \\
n^{0.126}  &\leq \frac{1}{0.20417813} \nonumber \\
n  &\leq 299,307 \nonumber \\
\end{alignat} 

Therefore, if the number of nodes is less than $2.99 \times 10^5$ nodes, every eigenvalue is guaranteed to lay in a unit circle. In other words, the non-linear dynamic system for the multi-hop star topology is locally stable at the equilibrium.
If there is a small perturbation around the equilibrium, the system is able to converge back to the equilibrium.

To prove the stability of other topologies, we can substitute $\Delta_i$ and $c_{i,j}$ in the Jacobian matrix with the value at the equilibrium of each topology. Then, finding the eigenvalues of the substituted Jacobian matrix. If we can bound that every eigenvalue lies in a unit circle, the algorithm is locally stable for such topologies. We conjecture that the proof of other topologies is similar to the proof of the star topology with the similar procedure.

\bibliographystyle{unsrt}  
\bibliography{mdwarf_bibfile}   

\end{document}